\documentstyle[floats,prd,aps]{revtex}

\input psfig.sty
\newcommand{\be}{\begin{equation}}
\newcommand{\ba}{\begin{eqnarray}}
\newcommand{\ee}{\end{equation}}
\newcommand{\ea}{\end{eqnarray}}

\begin{document}
\title{How the Universe Got its Spots}
\author{Janna Levin$^1$, Evan Scannapieco$^1$, Giancarlo de Gasperis${}^1$, 
Joseph Silk$^1$ and John D. Barrow${}^2$}
\address{
${}^{1}$Center for Particle Astrophysics, UC Berkeley}
\address{Berkeley, CA 94720-7304}
\address{$^{2}$Astronomy Centre, University of Sussex}
\address{Brighton BN1 9QJ, U.K.}
\maketitle

\begin{abstract}

The universe displays a three-dimensional pattern of hot and cold spots in
the radiation remnant from the big bang. The global geometry of the universe
can be revealed in the spatial distribution of these spots. 
In a topologically compact universe, distinctive patterns are especially
prominent in spatial correlations of the radiation temperature.
Whereas these patterns are usually washed out in statistical
averages, we propose a scheme which uses the universe's spots to observe
global geometry in a manner analogous to the use of multiple images of a
gravitationally lensed quasar to study the geometry of the lens.
To demonstrate how the geometry of space forms patterns,
we develop a simple real-space approximation to
estimate temperature correlations for any set of cosmological parameters 
and any global geometry.   
We present correlated spheres which clearly show topological pattern
formation for compact flat universes as well as for the compact
negatively curved space introduced by Weeks and another discovered by 
Best.
These examples illustrate how future satellite-based observations of the
microwave background can determine the full geometry of
the universe.

\end{abstract}


\vskip 10truept

\widetext

\twocolumn
\narrowtext
\begin{picture}(0,0)
\put(410,190){{ CfPA-98-TH-}}
\end{picture} \vspace*{-0.15 in}

>From the zebra's stripes to the leopard's spots, the animal kingdom displays
a diversity of coat patterns.
Following the innovative ideas of Turing 
\cite{turing}, mathematical biologists have posed and partly answered the
question of how the leopard got its spots. The fluctuation of enzymes
diffusing through the developing embryo can lead to the spatial pattern
formation displayed by animal coats. Both the geometry and size of the
animal exert a strong influence on differentiating patterns. For instance,
the broad cylindrical shape of the leopard's body favors spots while the
tapered tail induces stripes \cite{murray}. Remarkably, these diverse
features can arise from the properties of simple solutions to second-order
partial differential equations on the geometry and topology appropriate for
animal limbs, tails, or bodies. \footnote{
For example,  the two-dimensional Helmholtz
equation

\[
\nabla ^2\phi +k^2\phi =0,
\]
with $({\bf n\cdot \nabla })\phi =0$ on the boundary, has solutions which
describe different geometric tesselations of alternating regions with $\phi
>0$ and $\phi <0$, \cite{chris}$.$ For example, an infinite line of
one-dimensional stripes is described by the solutions $\phi =\cos kx,$ with $
k=n\pi $ for $n=\pm 1,\pm 2,..,$ whereas a solution $\phi =\frac 12[\cos
kx+\cos ky],$ with $k=\pi $ or $2\pi $ describes a chequer-board tesselation
of space with alternating  square spots, with $\phi >0$ and $\phi <0,$
inclined at $45\circ .$ }

Similarly, the global geometry of the universe can lead to distinctive
pattern formation by the normal modes of vibration of the universe in the
microwave sky, or even in the distribution of luminous galaxies. Unobscured
by complicated evolutionary effects, the cosmic microwave
background radiation (CMB) offers the best site to seek out
patterning. When light
last scattered off hot matter, small temperature fluctuations left
birthmarks in the radiation. These hot and cold spots in the primordial
radiation may be randomly distributed in an infinite universe, but if the
universe possesses a compact topology then distinctive patterns can be
generated. Such topologically induced pattern formation has already been
seen in simulated maps of the microwave sky \cite{{lbbs},{confus}}.

The diffusive processes in biological systems actually simplify an animal's
coat by singling out a particular mode and hence a particular pattern. By
contrast, the universe contains many competing modes. While some simple
large scale patterns are clearly evident in sky maps of the CMB temperature
fluctuations \cite{{lbbs},{confus}}, subtler patterns can be extracted from
the observations. In particular, a correlation in temperature between pairs
of points can scan the map and select out geometric features. Maps of
correlations for different topologies populate the cosmic zoo of
possibilities displayed in sections \ref{zoo} and \ref{zooh}.

CMB observations of the universe's hot and cold spots could be used as
real-space pictures of geometry. With correlated maps we could observe
topologically lensed images of the horizon at the time of last scattering.
The number and pattern of lensed images of the horizon provide information
about the geometry of the universe. A nice analogy is provided by
observations of gravitationally lensed images of a quasar, the pattern and
number of which reveal the geometry of the intervening lens. The universe's
spots are too small in angular scale to have been detected by the Cosmic
Background Explorer (COBE) but will be visible by the future satellite
missions, MAP (Microwave Anisotropy Probe) and {\it Planck Surveyor}.

The fluctuations in the CMB can be described by decomposing the relative
temperature fluctuations, $\delta T(\vec x)/T=\int d^3\vec k\hat \phi _{\vec 
k}\psi _{\vec k}$, into a series of eigenmodes $\psi _{\vec k}$ each of
which is a solution of Laplace's equation on the curved space of the
universe, $\left( \nabla ^2+k^2\right) \psi _{\vec k}=0$. The 
$\hat \phi _{ \vec k}$ are primordially seeded amplitudes that on average define a
spectrum of perturbations. Topology introduces boundary conditions which
create a discretized set of wave vectors and complex relations between the 
$\hat \phi _{\vec k}$ \cite{us}. This problem is tractable in the case of a
flat universe but, for a compact hyperbolic space, the identification rules
lead to such intricate boundary conditions on the modes that they cannot be
decomposed analytically \cite{chaos}. This is a symptom of the chaotic
hyperbolic flow of geodesics induced on these spaces. Microwave photons
moving on compact hyperbolic manifolds thus demand more indirect methods of
analysis. Cosmological observations favor a negative curvature and
mathematics favors compact hyperbolic manifolds by providing an infinite
number of candidates. The resistance encountered when subjecting these
manifolds to conventional eigenmode treatments has sparked recent interest
and produced several different approaches to the search for observational
probes of the universe's global topology. Three strategies have emerged: a
search for circles in the microwave sky \cite{css}, a search for
spatial pattern formation \cite{{lbbs},{us}}, and a direct attack via the
method of images \cite{bps}. All three approaches are necessarily related.
We will see aspects of each emerge in our zoo of topological examples.

For reviews and other interesting papers on cosmic topology see Refs. \cite
{{lum},{crystal},{ped},{quas}}.

\section{Topology and the angular correlation function}

Compact spaces are specified by ${\cal M}={\bf U}/\Gamma $ where ${\bf U}$
is the geometry of the infinite space before any compactifications are made. 
${\bf U}$ is also known as the universal covering space. The generators 
$\{g_k\}$ of the group $\Gamma $ provide the set of instructions for gluing
together points in ${\bf U}$ in order to render the space finite and
multiconnected. The $\Gamma $ forms a discrete subgroup of the full isometry
group of the covering space. A finite geometry has the same curvature as 
${\bf U}$. The universal covering space is usually assumed to be a space of
constant curvature, either the flat ${\bf E}^3$, the negatively curved 
${\bf H}^3,$ or the positively curved ${\bf S}^3$. The real universe does not have
perfectly constant curvature over the entire manifold: its true topologies
will then be a deformation of perfect polyhedra. The statement that the
universe looks homogeneous and isotropic on average indicates that constant
curvature will be a reasonable first assumption for finite manifolds which
are comparable to or smaller than the Hubble scale of the visible universe. We
limit ourselves to the subset of observationally viable manifolds with
covering spaces of either the flat ${\bf E}^3$ or hyperbolic ${\bf H}^3$
variety.

In order to scan the sky for evidence of compact topology we use the
correlation function for the temperature at two different points on the sky, 
\begin{equation}
\left\langle \frac{\delta T({\hat n})}T\frac{\delta T({\hat n}^{\prime })}T
\right\rangle =C({\hat n},{\hat n}^{\prime }),  \label{app}
\end{equation}
where ${\hat n}$ is a unit vector pointing from the Earth towards some
location on the sky. The sphere of radius $\Delta \eta {\hat n}$ defines the
surface from which we receive last-scattered photons. The size of the
radius, $\Delta \eta $, is the conformal time between the time of last
scattering and the present. The angular brackets denote an average over all
possible realizations. In a simply connected cosmos, the assumption that the
hot and cold spots are homogeneously and isotropically distributed ensures
that the correlation function depends on only one parameter, namely the
angle between the two points on the sky, that is $C({\hat n},{\hat n}
^{\prime })=C(\theta )$ where $\cos \theta ={\hat n}\cdot {\hat n}^{\prime }$. 
All of the information in an homogeneous and isotropic Gaussian field is
contained in $C(\theta )$.

By contrast, topological identifications always break isotropy (with the
exception of the projective space on ${\bf S}^3$) and most break homogeneity
as well. The hypertorus on ${\bf E}^3$ is the only homogeneous compact
space. Consequently, the correlation function, $C({\hat n},{\hat n}^{\prime
})$, depends fully on ${\hat n}$ and ${\hat n}^{\prime }$ and would require
four dimensions for a full representation. Predicting $C({\hat n},{\hat n}
^{\prime })$ is also more challenging on compact manifolds, particularly the
most unusual compact hyperbolic spaces. 

\begin{figure}[tbp]
\centerline{\psfig{file=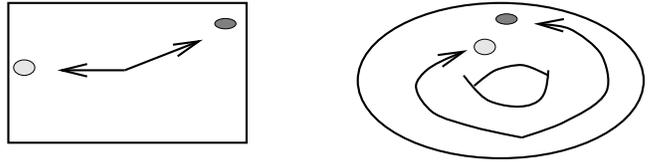,width=3.35in}} 
\vskip 5truept
\caption{Two points may appear to be far apart but if we identify opposite
sides of the rectangle we see that the same points are actually close
together on the compact torus. }
\label{close}
\end{figure}

We will introduce a real-space approximation which captures all the
important features necessary for examining global structure. Our
approximation is based on two physical observations. Firstly, two points
which appear to be widely separated may actually be close together, as in
Fig.\ \ref{close}: there is not a unique distance between points. Secondly,
the correlation between two points is strongly peaked around short
separations. While global topology drastically changes the large scale
perturbations, it does not have a strong impact
on scales much smaller than the
size of the physical space.
Therefore we expect the
correlation function between nearby points to be well approximated by the
correlation function of a simply connected universe.

Motivated by these two observations, we propose the following real space
approximation to the angular correlation function: we take the correlation
in temperature between two points on the surface of last scattering to be
the correlation function in a simply connected universe given their minimum
separation, that is, 
\begin{equation}
C_{{\cal M}}({\hat n},{\hat n}^{\prime })\approx C^U\left[ d_{{\rm min}}( 
\vec x({\hat n}),\vec x^{\prime }({\hat n}^{\prime }))\right] \ \ ,
\label{eq:approx}
\end{equation}
where $C^U$ is the correlation function on the universal cover, ${\vec x}({ 
\hat n})$ is the physical location of a point on the surface of last
scattering, and $d_{{\rm min}}$ is the minimum distance between the two
points in the topological space. To find the minimum distance we locate the
image points with the generators of the identifications. The point $\vec x 
^{\prime }$ has first neighbor images at locations $\vec y_k=g_k\vec x 
^{\prime }$ and second neighbor images at locations $\vec y 
_{k_2k_1}=g_{k_2}g_{k_1}\vec x^{\prime },$ etc. More concisely, the images
out to order $m$ can be written as 
\begin{equation}
\vec y_{k_m,..,k_1}=\prod_i^mg_{k_i}\vec x^{\prime }(\hat n^{\prime })\ \ .
\end{equation}
The image point which lands closest to $\vec x(\hat n)$ determines $d_{{\rm  
min}}$. Note that all of the effects of topology appear only in the
minimization of the separation between the two points. Thus, the correlation
function can be approximated using only a knowledge of the identifications
that compactify the space: the very relations that are used to specify the
topology.

Our approximation is closely related to the method of images employed in
Refs. \cite{bps}. The correlation function computed by the method of images
is a sum over {\it all} of the copies of the image points in the simply
connected space with the same curvature, 
\begin{eqnarray}
C(\hat n,\hat n^{\prime })=\lim_{r_{*}\rightarrow \infty } &{}&\sum_{\gamma
_i}C^U(d({\hat n},\gamma _i{\hat n}^{\prime }))  \nonumber \\
&&-{\frac{4\pi }V}\int_0^{r_{*}}dr\sinh ^2rC^U(r)\ \ ,
\end{eqnarray}
where the second term is a regularizer introduced to control the exponential
proliferation of images, particularly on ${\bf H}^3$, as the radius $r_{*}$
increases relative to the volume of the fundamental domain, $V$ \cite{bps}.
The $\gamma $ are composite elements of the group $\Gamma $. Our
approximation amounts to keeping only the dominant term in the sum over
images. 
Maintaining the first term is by far 
less cumbersome than summing an infinite number
of terms and fares well in approximating the exact correlation function.
The approximation is also quite valuable for our purposes since
we are able to easily
include the essential effects of the smoothing across the
horizon at the time of last scattering and the relevant microphysics at work
over small separations. 
The small-scale physics at the time of last
scattering is important for distinguishing topologically lensed images from
fictitious correlations as explained in \S \ref{obsp}.  The physical
processes operating at last scattering can in principle 
be folded into the method of images to refine our approximation  
and sharpen the focus on our
pictures of geometry. 
The thickness and velocity
of the surface of last scatter will also induce additional corrections.
A test of our approximation when applied to the flat
geometries is given in detail in \S \ref{obsp}.

Given an estimate of the full four-dimensional $C_{{\cal M}}(\hat n,\hat n 
^{\prime })$, we could build realizations of a map of $\delta T(\hat n)/T$
for any topology, even without the eigenmodes. A given universe would be
obtained by a random realization of a Gaussian-distributed variable with a
mean $\left\langle \delta T(\hat n)/T\right\rangle =0$ and variance $C_{ 
{\cal M}}(\hat n,\hat n^{\prime })$. However, we can do better. 
As $C_{{\cal M}}({\hat n},{\hat n}^{\prime })$ is a function on two copies
of the sky, it necessarily must be calculated numerically for $N_{pix}^2$
values, where $N_{pix}$ is the number of pixels into which the sky is
divided. This is cumbersome, if not practically impossible, to carry out for
many topologies on a reasonably finely grided sky map, particularly if we
aim at the resolutions expected of future CMB satellite missions.
Furthermore, many of the peaks of this function occur at values where ${\hat 
n}$ and ${\hat n}^{\prime }$ are separated by a small angle on the sky.
Since these points would be close together regardless of global topology,
much of the base space of $C_{{\cal M}}({\hat n},{\hat n}^{\prime })$ is
useless in discriminating between topological properties of cosmological
models.

Instead of building full $\delta T(\hat n)/T$ maps, we build selective
correlation maps. An illustrative example is the antipodal map constructed
by evaluating the correlation function at antipodal points along the sky, 
\begin{equation}
{\rm A}({\hat n})\equiv C_{{\cal M}}(\hat n,-{\hat n}).
\end{equation}
In a simply-connected universe ${\rm A}(\hat n)$ would produce on average
nothing more than an overall monopole. Fortuitous correlations may appear at
random in a given realization but there would be no defining structure. 
In a universe with compact topology a great deal of structure can be
surveyed by means of this one simple correlation. While the best analysis
for any particular topology will make use of the full correlation function,
the antipodal map will prove to be a most useful survey tool. To get a feel
for the way in which correlations reflect global topology, we make a gallery
of correlated spheres first for all flat topologies and then for a few
choice compact hyperbolic spaces. We also consider correlations under
symmetries other than antipodal as the specific space demands. In order to
represent the correlations as $2D$ maps we consider correlations of the form 
\begin{equation}
C_g(\hat n)\equiv C_{{\cal M}}(\hat n,g\hat n/||g\hat n||)\ \ ,
\end{equation}
where again $g$ is an element of the group $\Gamma $.
For the compact hyperbolic spaces, we also correlate an arbitrary
point with the rest of the surface of last scatter.
In many ways these point-to-sphere correlations, are the most
promising.  They very dramatically reveal geometric patterns and 
they do not require any foreknowledge of the symmetries of the space.

\section{The Cosmic Zoo}

\label{zoo}

\subsection{Flat Topologies}

Compact flat spaces have already fallen out of favor as small universes. The
first suspect for investigation was the simplest hypertorus \cite{flatg}.
The fundamental domain is a parallelepiped with opposite faces identified in
pairs. As shown by Stevens, Scott and Silk \cite{sss}, the square hypertorus
suffers a sharp truncation of long wavelength power in temperature
fluctuations, too sharp to be consistent with COBE observations unless the
box is larger than about $40\%$ of the observable universe. Later,
anisotropic tori were studied in Refs. \cite{smoot} using symmetry methods,
and were similarly bound.  The
tightest limits on an equal-sided square hypertorus
were obtained in Ref. \cite{bps} using the method of
images, placing the topology scale just beyond the observable universe with $
h\ge 2.19\Delta \eta $ where $h$ is the length of the side of the square and $
\Delta \eta $ is the conformal time since last scattering. 

\begin{figure}[tbp]
\centerline{{\psfig{file=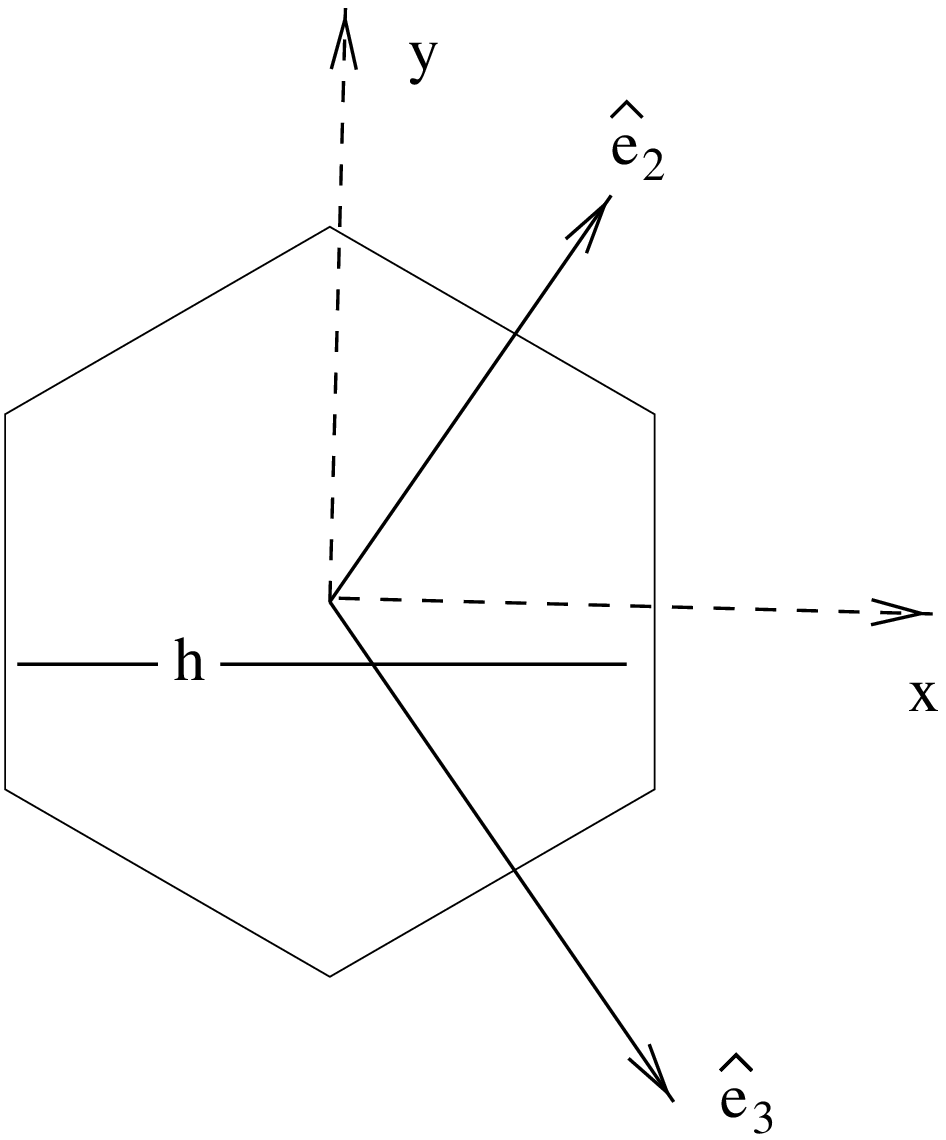,width=1.2in}} \hskip 5truept { 
\psfig{file=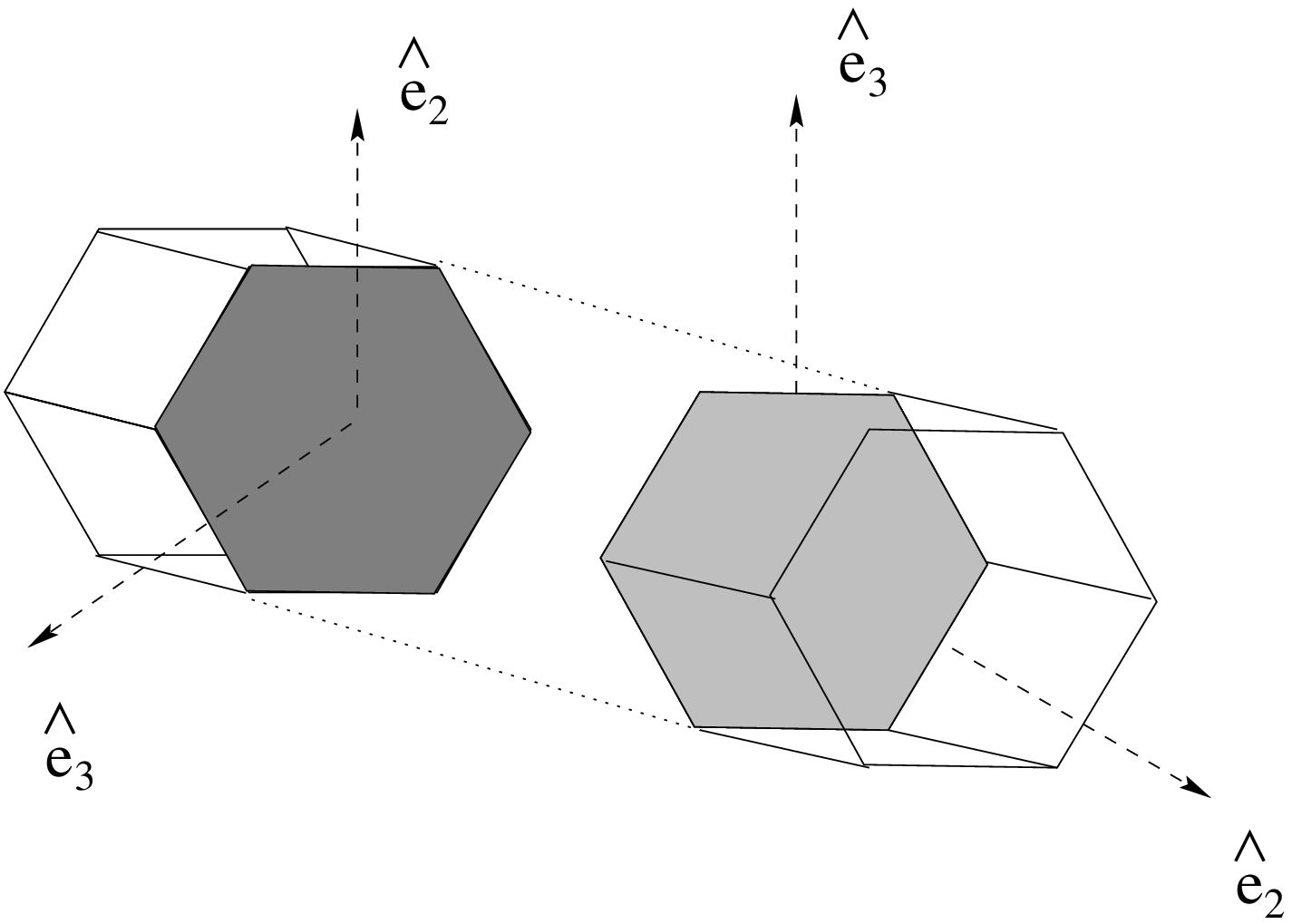,width=2.25in}}} \vskip 5truept
\caption{The hexagonal geometry. The prism face is glued with a twist of $
2\pi/3$ to create a topology distinct from the hypertorus. }
\label{hexgeom}
\end{figure}

\begin{figure}[tbp]
\centerline{\ {\psfig{file=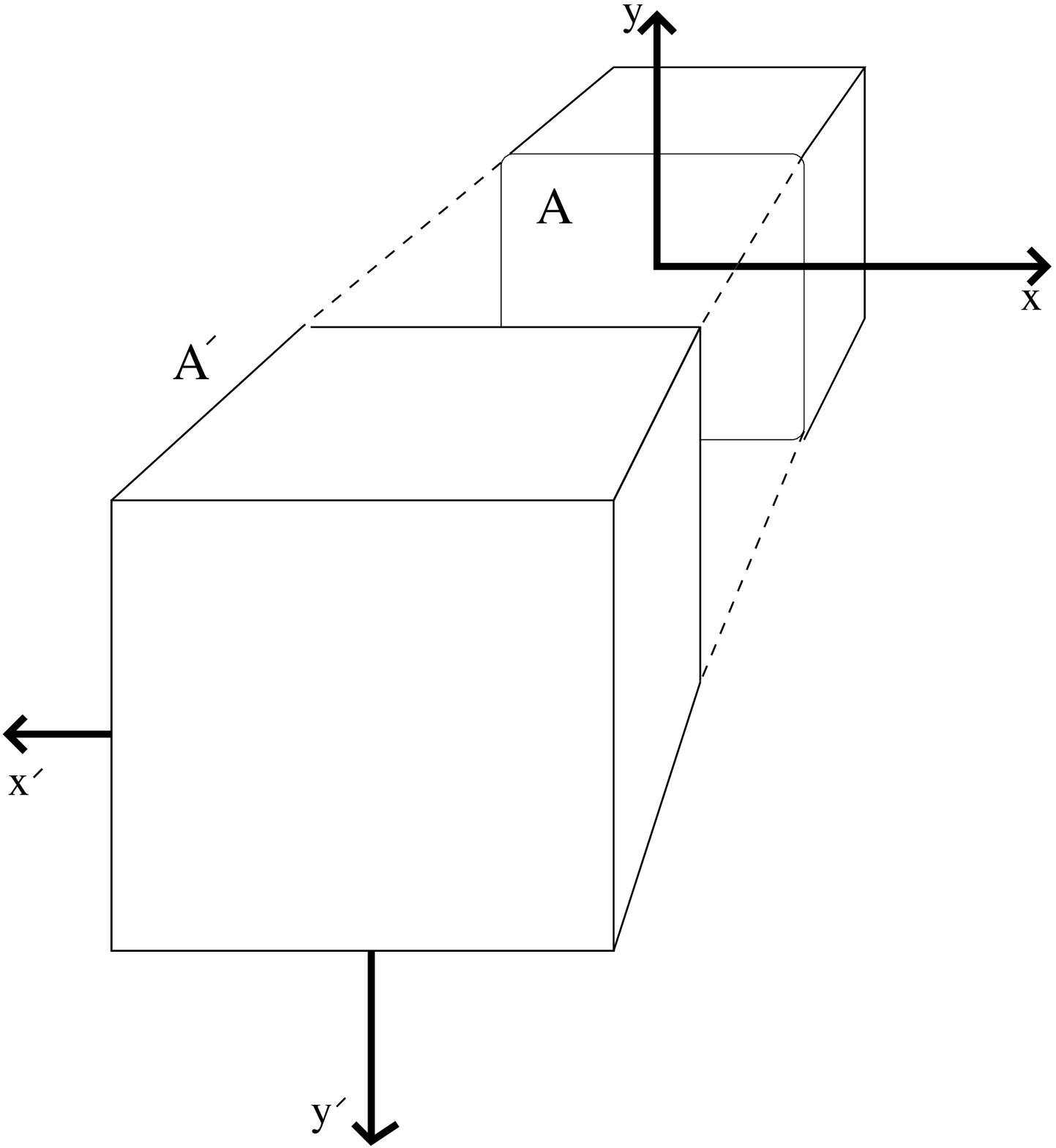,width=1.75in}} { 
\psfig{file=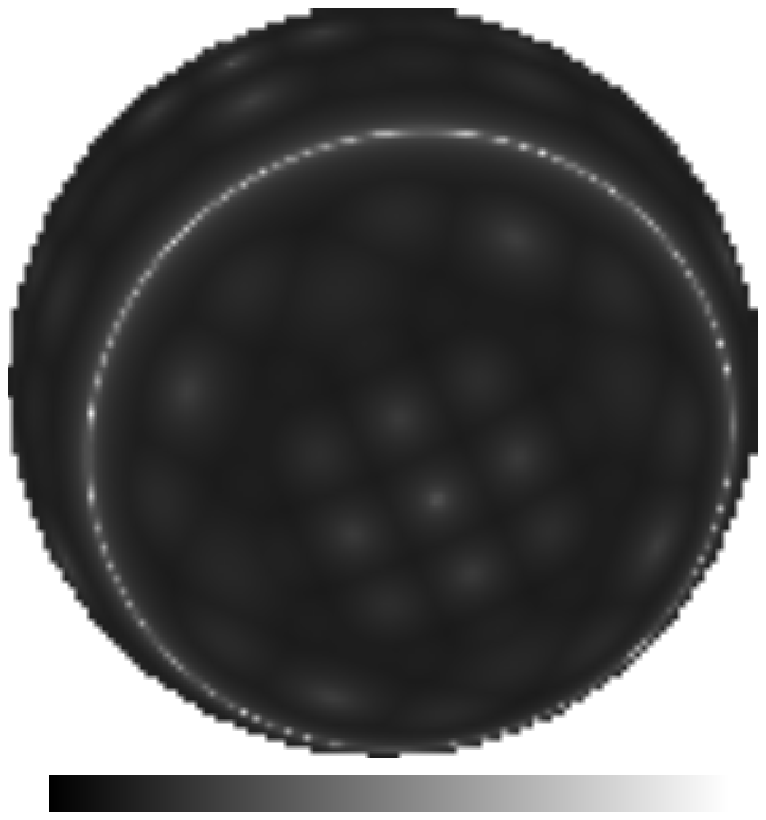,width=2in}} } \vskip 15truept
\caption{${\rm A({\hat n})} $ for a $\pi/2$-twisted space. }
\label{pi2pat}
\end{figure}

With the universal covering space of ${\bf E}^3$, there are only five more
orientable, compact spaces that can be constructed \cite{wolf}. Three are
built by identifying the faces of a parallelepiped with relative twist (see
Fig.\ \ref{pi2pat}), and two are constructed by identifying the faces of a
hexagonal prism with the prism faces twisted by $2\pi /3$ or $\pi /3$
relative to each other before being identified. In Ref. \cite{{us},{us2}},
we derived the eigenmode spectrum explicitly for all twisted cases and
showed that the angle-averaged power spectrum is incompatible with COBE for
equilateral spaces unless the universe was really big, if not actually
infinite (i.e. topological identification scale $\ge 80\%\Delta \eta $). The
compact flat spaces are thus of limited interest, although polyhedra with
sides of disparate lengths may still be viable.

However, since we do have eigenmodes for all flat topologies \cite{us},
these topologies are important testing grounds for any method which attempts
to circumvent the eigenvalue problem. We therefore use ${\bf E}^3$ to test
our real space approximation before moving on to compact topologies of ${\bf  
H}^3$.

For illustration, we spend time explaining the features in the markings for
a $2\pi /3$-twisted hexagonal prism. The manifold is ${\bf E}^3/\Gamma $
with $\Gamma $ the group of instructions for identifying the faces of the
hexagon. 
Define a coordinate system for which $\hat e_1$ is orthogonal to the
hexagonal face and the non-orthogonal vectors $\hat e_2,\hat e_3$ span the
face as in Fig.\ \ref{hexgeom}. $\Gamma $ is generated by $\{g_1,g_2,g_3\}$
where $g_1$ generates a rotation through $2\pi /3$ about the $\hat e_1$ axis
combined with a translation orthogonal to the hexagonal face through a
distance $c$, $g_2$ effects a translation along $\hat e_2$ through $h,$ and $
g_3$ effects a translation along $\hat e_3$ also through $h$. Another way to
visualize ${\bf E}^3/\Gamma $ is to glue copies of the fundamental domain
together according to the identification rules. In this way ${\bf E}^3$ can
be completely tiled with layers of hexagons separated by the length of the
prism direction.

\begin{figure}[tbp]
\centerline{\psfig{file=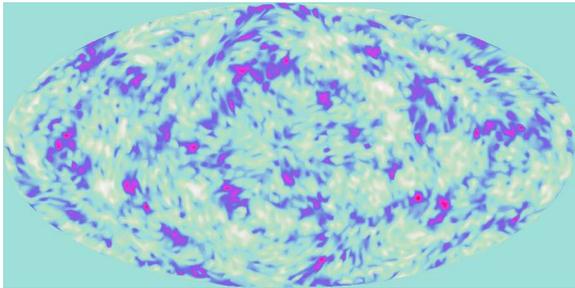,width=3in}}
\caption{ A typical map of $\delta T/T(\hat n)$ in a hexagonal prism with $
h=0.8\Delta \eta$. }
\label{hexmap}
\end{figure}

We can use these symmetries to build the discrete spectrum of eigenmodes
which describe fluctuations on this space as was done in Ref. \cite{us}.
Using these modes a typical map of $\delta T(\hat n)/T$ for a simulated
compact hexagonal universe can be created. We show this for comparison in
Fig.\ \ref{hexmap}. There is something at work in the map of ${\delta T({ 
\hat n})/T}$ but it is hard to define.

\begin{figure}[tbp]
\centerline{\psfig{file=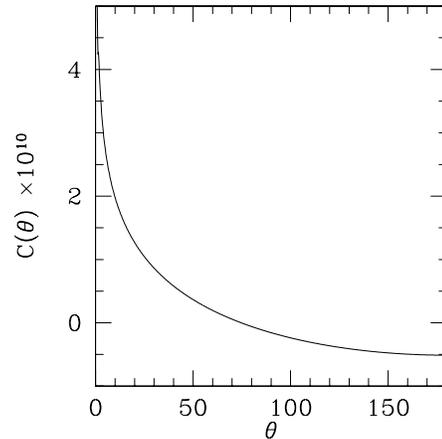,width=2.4in}}
\caption{$C^U(\theta)$ for a flat, COBE normalized CDM cosmology. Note that
the becomes negative due to the definition of $\delta T/T$.}
\label{ctheta}
\end{figure}

If, however, we inspect the idealized zero-variance antipodal map, the
pattern jumps out at us. These antipodal maps were built with our
approximation and {\it did not require the full mode solutions}. We locate a
point on the sphere of last scattering (SLS) $\vec x=\Delta \eta \hat n$ and
compare the correlation of that temperature with its antipodal point $\vec x 
^{\prime }=-\Delta \eta \hat n^{\prime }$. If the universe were simply ${\bf  
E}^3$, that is, flat and unconnected, then these points should be totally
uncorrelated on average. Given that this space is multiconnected, the two
opposed points may in fact be close together as demonstrated for the $2D$
surface of Fig.\ \ref{close}. We approximate the correlation between
antipodal points as the correlation that the points would have in a simply
connected space given their minimum separation as in (\ref{app}). Antipody
is then approximated as 
\begin{equation}
\left\langle {\rm A({\hat n})}\right\rangle \simeq C^U\left[ d_{{\rm min}}( 
\vec x(\hat n),\vec x^{\prime }(-\hat n))\right]
\end{equation}
where $C^U(\theta )$ is obtained from CMBFAST \cite{cmbfast} for a standard,
flat CDM cosmology ($\Omega _b=.05,\Omega _c=.95,H_0=50Kms^{-1}Mpc^{-1}$)
with the dipole component calculated as per a flat power spectrum. This is
illustrated in Fig.\ \ref{ctheta}. We minimize the distance by first taking
the image of these points under the action of the generators of the group $
\Gamma $ until we relocate them within the fundamental domain. 
Leaving one of these relocated points fixed, we consider all images of the
second point that lie within one of the nearest neighbors of the fundamental
domain, and choose $d_{{\rm min}}$ as the shortest distance from the first
point to one of these images. Note that only by considering all of these
images, including those that are diagonally located relative to the
fundamental domain, can we be sure that our definition of distance depends
on the overall topology of the space rather than on the particular
coordinates that we use to fix the fundamental domain. This method results
in an overall monopole component of the antipodal map which we wish to
discard. 
We simply remove the monopole as calculated from the antipodal map alone,
with the understanding that this is to be compared with a measured map that
is similarly normalized.

\begin{figure}[tbp]
\centerline{\quad\quad\quad\quad\quad\quad  
\psfig{file=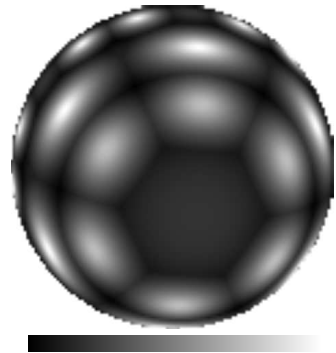,width=2.25in}} \vskip 15truept
\caption{Orthographic projection of ${\rm A({\hat n})}$ at a resolution 
of $20$ arcminutes for a hexagonal
prism with $h = b = c = 0.8 \Delta \eta$. The observer is at the origin. }
\label{hexpat}
\end{figure}

It is customary to use the Aitoff projection (as in Fig.\ \ref{hexmap}) to
view the map of $\delta T({\hat n})/T$ so as to see the entire sky. For the
antipodal map we prefer the orthographic projection which shows the genuine
shape of the surface of last scattering. In this three-dimensional view,
there is no distortion of the sky pattern. No information is lost by
limiting ourselves to a view of only half the sky at a time since ${\rm A({ 
\hat n})}$ is by definition symmetric under $\pi $.

\begin{figure}[tbp]
\centerline{{\psfig{file=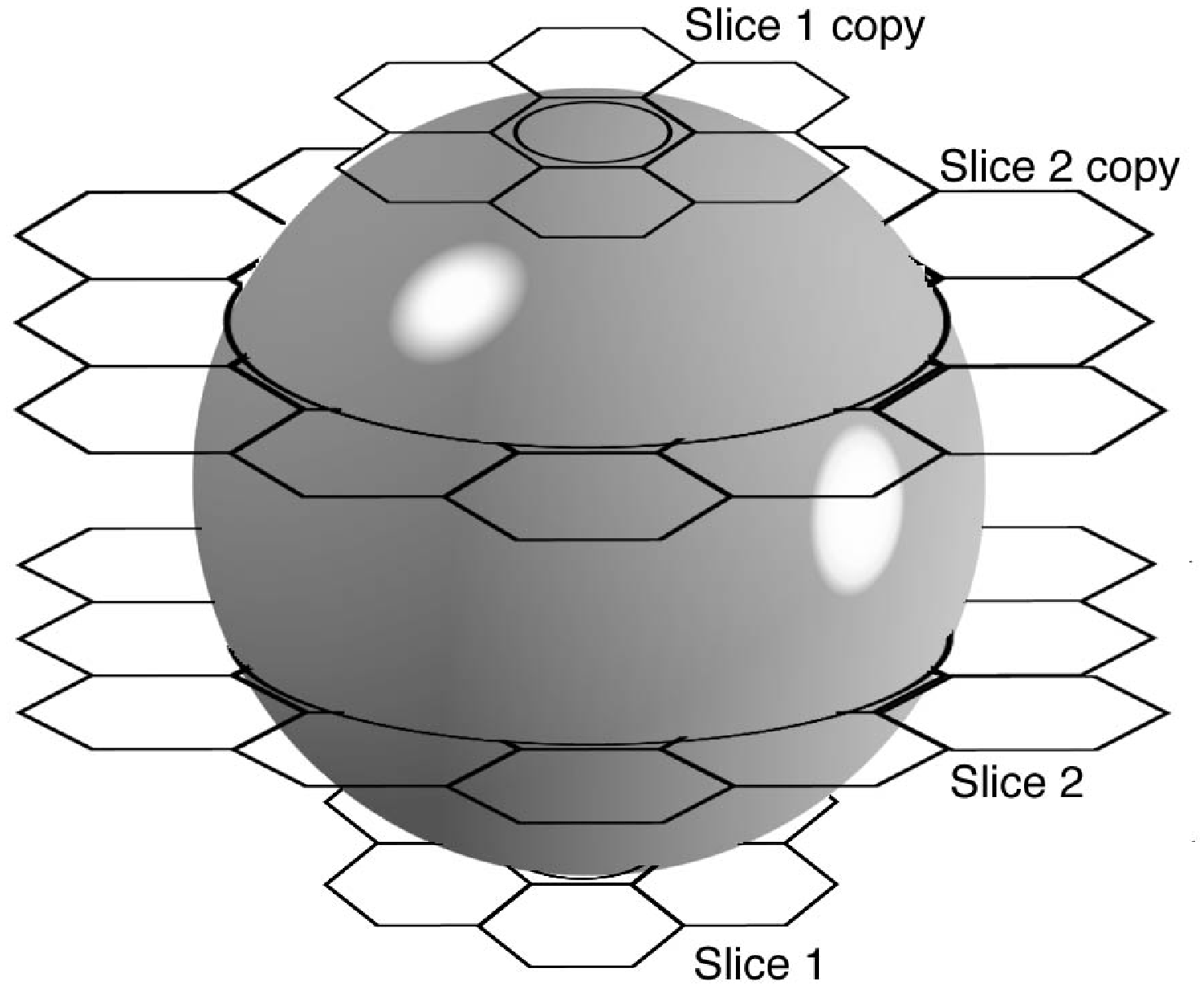,width=3in}}} \centerline{{ 
\psfig{file=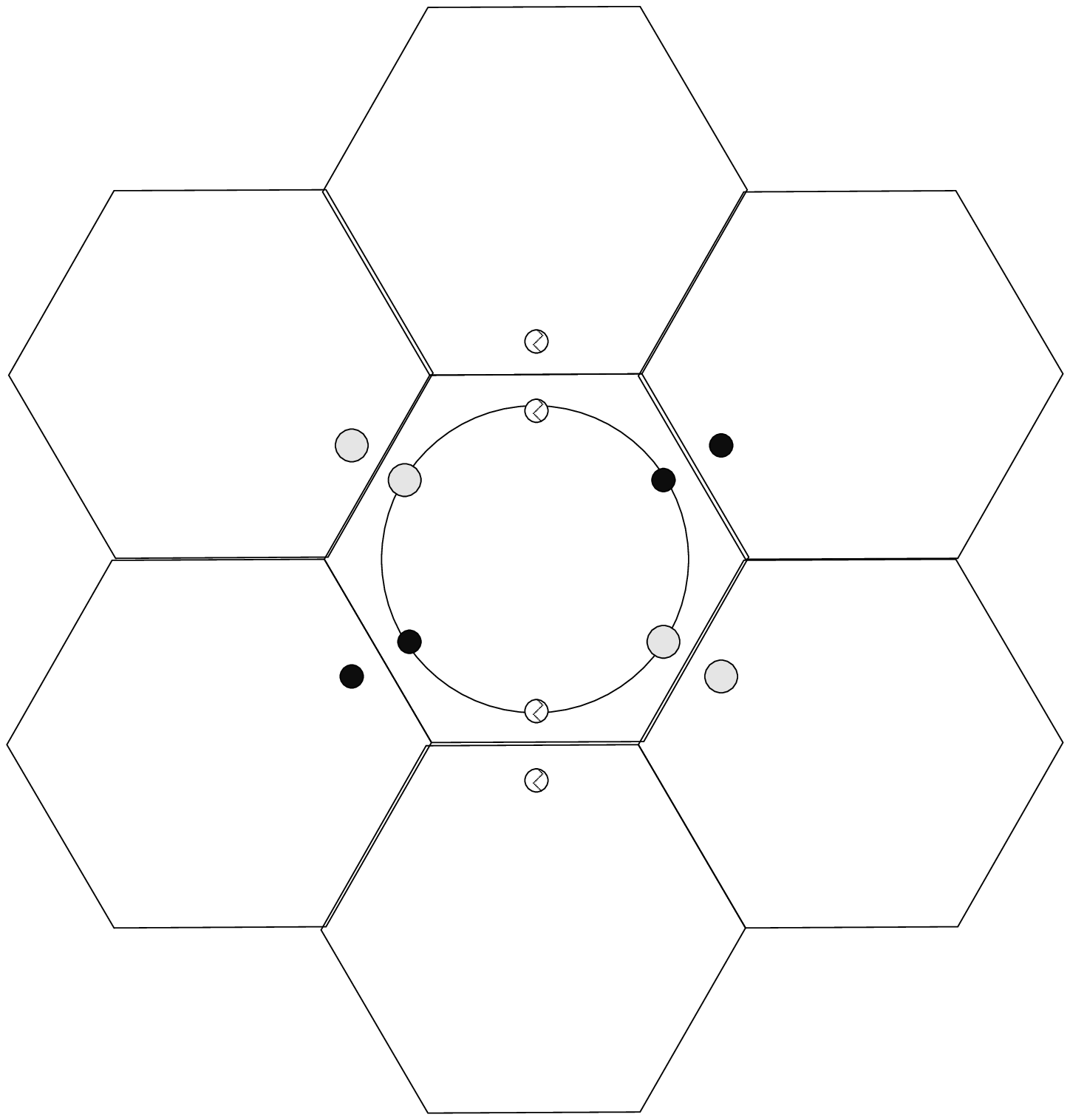,width=1.25in}} { 
\psfig{file=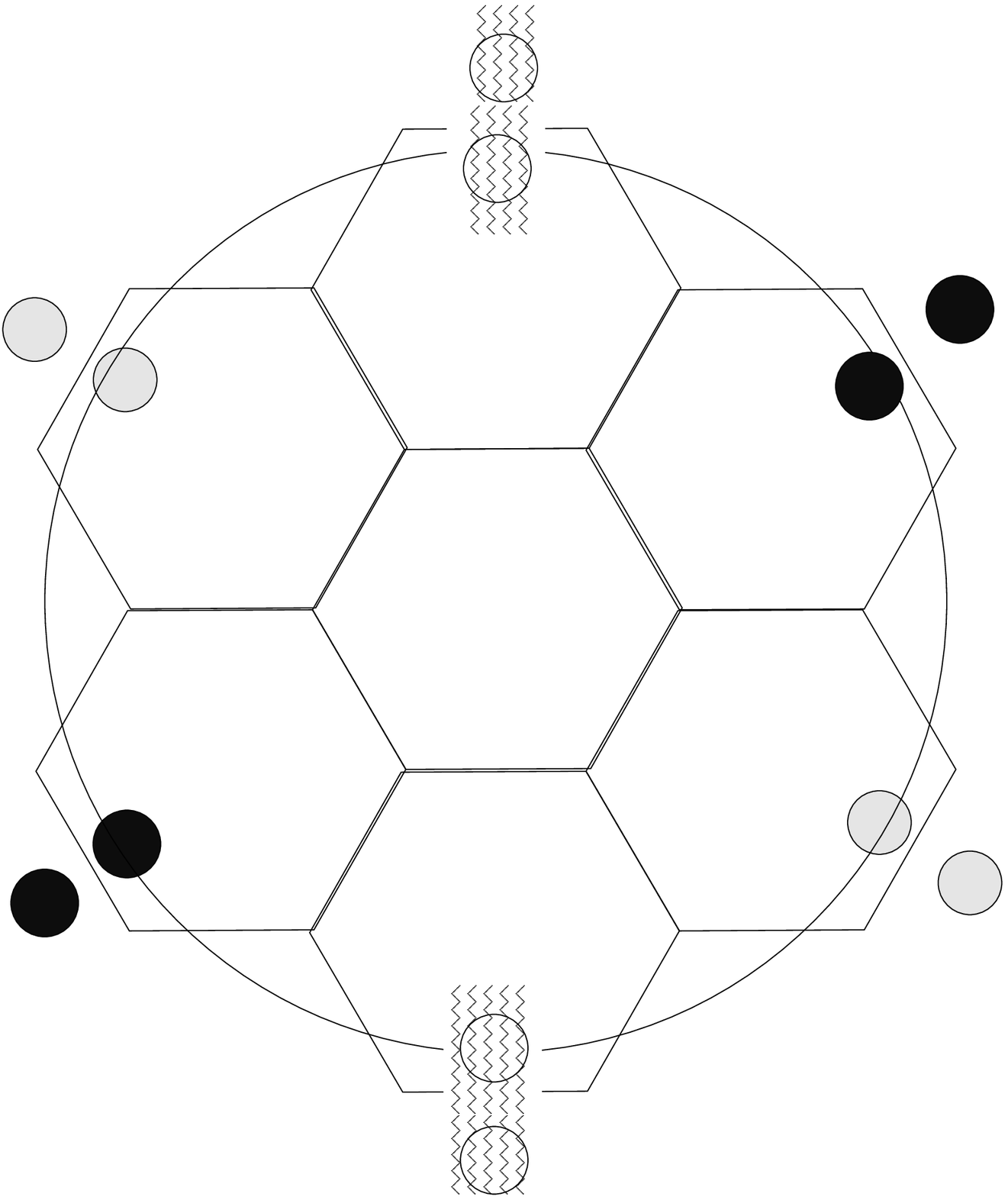,width=1.25in}}} \vskip 5truept
\caption{The surface of last scatter intersects the layered tiling of flat
space. Each full tile represents a copy of the fundamental domain. Slice 1
is represented on the lower left and slice 2 on the lower right. The dots
show correlated points picked up in the antipodal map and explain the
emergence of the hexagonal geometry in Fig.\ \ref{hexpat} }
\label{slices}
\end{figure}

There is clearly a hexagon in the antipodal map. The intersection of the
hexagonal layers with the spherical surface of last scattering defines rings
centered on the prism direction and are clearly picked out by ${\rm A({\hat n 
})}$. The rings of correlated spots occur since the correlation between
points on the sphere of last scattering separated by $\pi $ for this
geometry is the same as the correlation between opposites sides on a
circular slice through the sphere taken along the hexagonal plane, as
demonstrated in Fig.\ \ref{slices}. 
The bright primary spots are identical points
at the core and are just correlated away from the center.
In addition to these identical images, there are also secondary
spots picked up due to the correlation of 
regions which just near each other.

These rings of structure
are not the circles of the sky of Ref. \cite{css}, although they
are related. 
As pointed out
there, pairs of identical circles occur in the microwave sky due to the
intersection of the surface of last scattering with copies of itself. 
None of the
circles are located by ${\rm A({\hat n})}$ for the $2\pi /3$ twist because
none of the circles in this space are paired under a $\pi $-symmetry.
For rings separated along the prism direction by multiples of $3c$, the
hexagonal faces have completed a full rotation and antipody compares one
point to its opposite face. 
We continue to call
the concentric collections of spots in the correlated $A(\hat n)$ maps
`rings'
and we reserve `circles' for intersections of the copies of the surface of
last scattering with itself. 
So, the primary
spots on a given ring lie on
different pairs of circles in the sky.
The rings of secondary spots do not lie on circle pairs.

\begin{figure}[tbp]
\centerline{\quad\quad\quad\quad\quad\quad  
\psfig{file=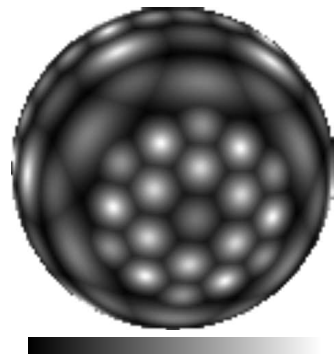,width=2.25in}} \vskip 15truept
\caption{Orthographic projection of ${\rm A({\hat n})}$ for a hexagonal
prism with $h = b = c = 0.6 \Delta \eta$. }
\label{hexpat2}
\end{figure}

As the topology scale gets smaller, there are more rings of patterns and the
first ring appears closer to the caps as shown in Fig.\ \ref{hexpat2}.

The size of a spot will be set by the Silk damping which smooths
fluctuations on small scales. For separations which exceed this length, we
expect correlations quickly to die off. Since the Silk damping scale is
smaller than the horizon size at the time of decoupling we expect the
angular size of these spots to be too small for COBE to have detected. The
beam smearing in the COBE experiment would dilute a spot over such a large
angle that these bright markings would be below the signal to noise
sensitivity of the detector. However, any experiment with high enough
resolution and sensitivity to probe the Doppler peaks will be able to
resolve these pattern on the sky. The planned missions MAP and {\it Planck
Surveyor} would thus be able to detect the universe's spots.

\begin{figure}[tbp]
\centerline{\quad\quad\quad\quad\quad\quad { 
\psfig{file=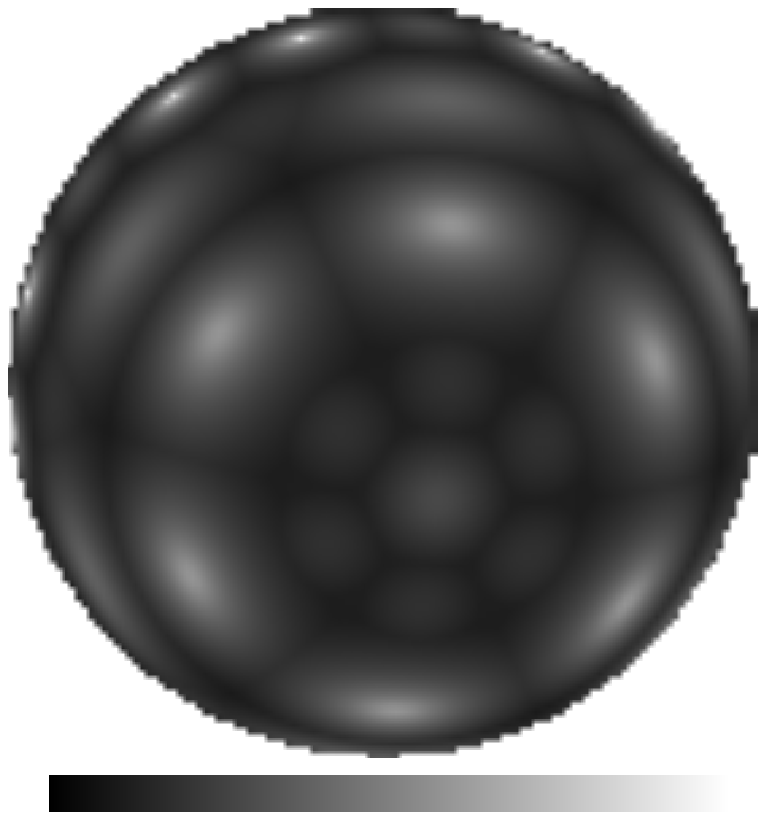,width=2in}} 
\psfig{file=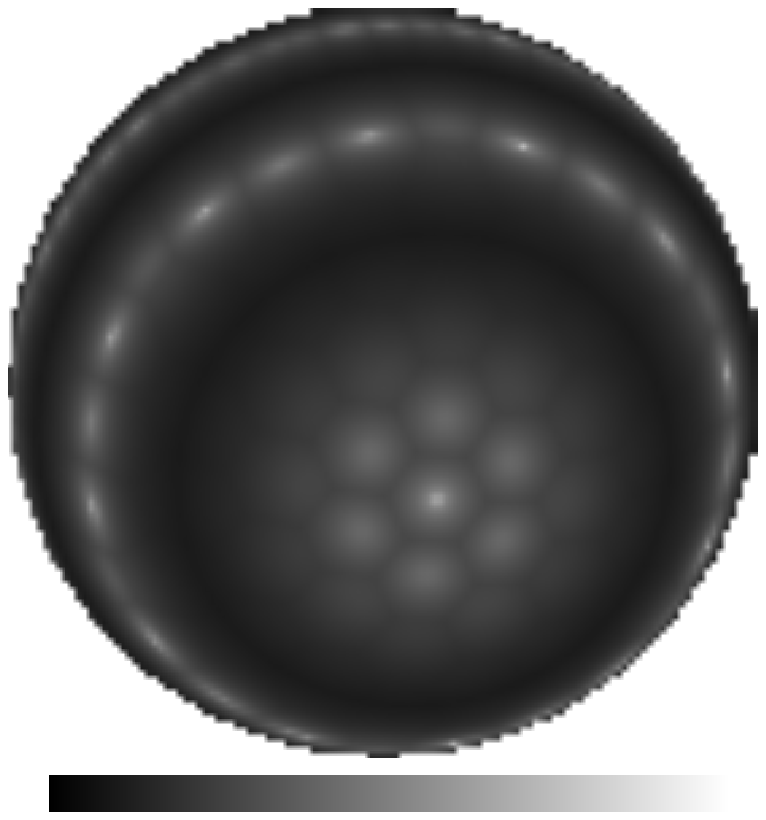,width=2in}} \vskip 15truept
\caption{${\rm A({\hat n})} $ for a hexagonal prism with a $\pi/3$ twist $h
= c = 0.75 \Delta \eta$ there are no circles. To the right the $z$ direction
is $.24$ while $h=1$. There are circles }
\label{hex2}
\end{figure}

The other topology built from the hexagonal plane tiling involves a relative
twist of the prism face through $\pi /3$. As seen in the left-most panel of
Fig.\ \ref{hex2}, there are again correlated spots on rings. When we make
the space sufficiently small so that at least three copies of the
fundamental domain fit within the SLS the first pair of circles appears
in $A(\hat n)$, as
shown in the right-most panel of Fig.\ \ref{hex2}. 

\begin{figure}[tbp]
\centerline{\ \quad\quad\quad\quad\quad\quad { 
\psfig{file=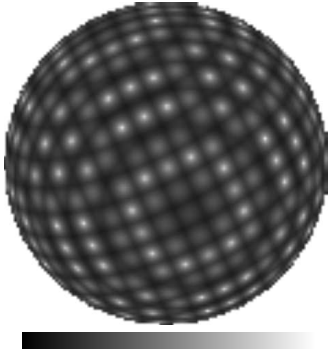,width=2.25in}} } 
\vskip 15truept
\caption{${\rm A({\hat n})} $ for the torus with $h = b =c = 0.31 \Delta
\eta $. }
\label{torpat}
\end{figure}

The other four topologies are built from a parallelepiped as the fundamental
domain, as for the simplest hypertorus. No circles will be located by ${\rm  
A({\hat n})}$ for the hypertorus, which is not to say that the hypertorus
has no circles, just that none of the pairs of circles are rotated by $\pi $
relative to each other. Since the hypertorus does not involve any rotations
of the faces, all three directions are picked out equally by antipodal
pairings, as seen in the small torus of Fig.\ \ref{torpat}. The square
geometry of the fundamental domain is blatantly traced out by the correlated
spots leading to the disco ball effect.

For the twisted parallelepipeds, ${\rm A({\hat n})}$ will locate the axis of
symmetry along which the faces are twisted. The correlated spots still trace
out the symmetric square of the equilateral untwisted directions, as shown
in Fig.\ \ref{pi2pat}. For the $\pi /2$ twisted space, the circles appear in 
${\rm A({\hat n})}$ if more than two copies of the fundamental domain fit
inside the observable universe. At least one pair of circles will appear
always for the $\pi $-twisted space as well as for the last compact topology
built by gluing a parallelepiped by a series of diagonal translations and $
\pi $ twists. 

The menagerie in this flat zoo is a testament to the influence of geometric
patterns on the universe's markings. An antipodal correlation is just one
statistic but note how clearly antipodal correlations reveal the other
symmetries of the space. For instance, the prism direction perpendicular to
the hexagonal face is clearly identified. Once the symmetries of the space
do begin to become apparent, the future data on the CMB can be
systematically scanned for correlations under other symmetries of a given
topological space: $C_{{\cal M}}({\hat n},g{\hat n}/||g{\hat n}||)$. In this
way the entire fundamental group $\Gamma $ can be isolated, thereby
determining the full geometry of the space, as was also suggested in Ref. 
\cite{weeks}.

\begin{figure}[tbp]
\centerline{\quad\quad\quad\quad\quad\quad  
\psfig{file=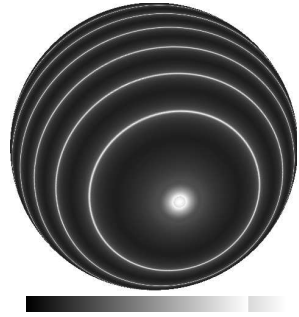,width=2.in}} \vskip 15truept
\caption{The sphere of correlations $C(\hat n(x,y,z),\hat n(x,y,-z))$ for
the hypertorus. The topology scale is $0.4\Delta \eta$. }
\label{hypz}
\end{figure}

The indicator ${\rm A({\hat n})}$ is particularly convenient because it can
be visualized on an half-sphere without the specification 
of a particular axis,
but once an axis of symmetry is chosen, correlations at other angular
separations can similarly be examined. As an example of such a correlation,
we consider $C_{{\cal M}}(\hat n,\hat n^{\prime })$ where we merely compare
the point $(x,y,z)$ to the point $(x,y,-z)$ for the hypertorus of size $0.4$
of the SLS. We assume that this axis has been identified by examining the
overall antipodal properties of the sky. We expect this correlation to pick
up circles and it does, as demonstrated in the Fig.\ \ref{hypz}.

\begin{figure}[tbp]
\centerline{\quad\quad\quad\quad\quad\quad  
\psfig{file=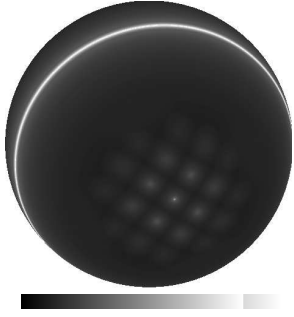,width=2.in}} \vskip 15truept
\caption{The correlated sphere comparing points on the surface of last
scatter with points related by a $\pi/2$-twist around the $z$-axis. The
space is a $\pi/2$-twisted square with sides of length $0.4\Delta\eta $.}
\label{pix}
\end{figure}

As another example we compare, for the $\pi /2$-twisted space of size $0.4$
the radius of the SLS, the correlation between two points separated by a
rotation of $\pi /2$ around the $x$ axis. In Fig.\ \ref{pix} we display the
result of applying the correlation function 
\begin{equation}
C_{{\cal M}}\left[ d_{{\rm min}}(\vec x(\theta ,\phi ),\vec x^{\prime
}(\theta ,\phi +\pi /2))\right] \ \ .
\end{equation}

\subsection{Observing the Universe's spots}

\label{obsp}

Our topological zoo of maps
makes a compelling visual argument for a pattern-driven approach to the
search for cosmic topology. Nevertheless, they are still only
ensemble-averaged images calculated using an approximation method. As the
real microwave sky contains no such averages or approximations, 
we are left wondering if we should
believe our eyes.
We must look
more closely at what is observed.

In this section, we address the accuracy of our method by comparing our
approximate maps with exactly calculated ensemble-averaged maps as well as
with simulated realizations of the microwave sky. We consider the flat
hypertorus which has a simple eigenmode decomposition of the temperature
fluctuations, 
\begin{equation}
{\frac{\delta T}T}(\hat n)\propto \sum_{\vec k}\hat \phi _{\vec k}\exp
\left( i\Delta \eta \vec k\cdot {\hat n}\right) ,  \label{eq1}
\end{equation}
with $\vec k=2\pi (n_x/h,n_y/b,n_z/c)$. The $\hat \phi _{\vec k}$ are
primordially seeded Gaussian amplitudes that obey the reality condition $
\hat \phi _{\vec k}=\hat \phi _{-\vec k}^{*}$ and as an ensemble define the
spectrum 
\begin{equation}
\left\langle \hat \phi _{\vec k}\hat \phi _{\vec k^{\prime
}}^{*}\right\rangle ={\frac{2\pi ^2}{k^3}}{\cal P}(k)\delta _{\vec k,\vec k 
^{\prime }}\ \ .
\end{equation}
With this decomposition we can construct the correlation function between
any two points on the sky as 
\begin{eqnarray}
&&C(\hat n,\hat n^{\prime })\,=\,\left\langle {\frac{\delta T}T}(\hat n){ 
\frac{\delta T}T}(\hat n^{\prime })\right\rangle \  \\
&\propto &\sum_{\vec k}{\frac{{\cal P}(k)}{k^3}}\exp \left( i\Delta \eta  
\vec k\cdot (\hat n-\hat n^{\prime })\right) .  \nonumber  \label{eq:cnn}
\end{eqnarray}
The antipodal correlation on the hypertorus is the simple case, $C(\hat n,- 
\hat n)$, and 
\begin{equation}
\left\langle {\rm A({\hat n})}\right\rangle \propto \sum_{\vec k}{\frac{ 
{\cal P}(k)}{k^3}}\exp (i2\Delta \eta \vec k\cdot \hat n).  \label{eq2}
\end{equation}
up to an overall normalization. This is nothing more than a Fourier
transform of the power spectrum, and can be computed exactly.

\begin{figure}[tbp]
\centerline{\quad\quad\quad\quad\quad\quad  
\psfig{file=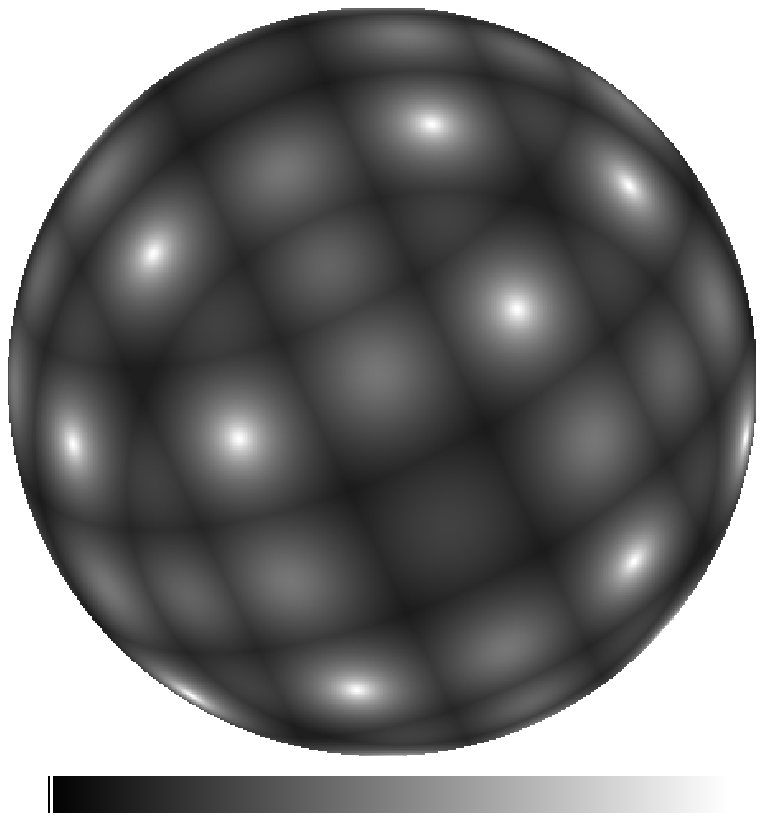,width=2.in} 
\psfig{file=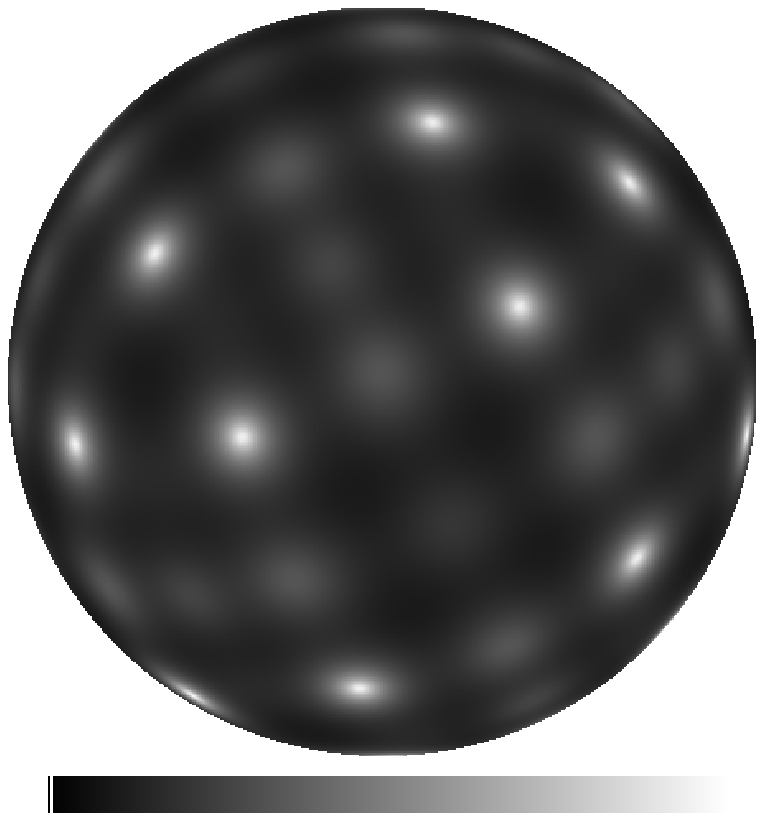,width=2.in}} \vskip 15truept
\caption{A comparison of the real-space approximated antipodal map (left)
with the exact ensembled-averaged map (right) for a torus space with $h = b
= c = .8 \Delta \eta$.}
\label{checkappx}
\end{figure}

In figure \ref{checkappx} we compare the results of our approximation with
the exact eigenmode decomposition of eqn.\ (\ref{eq2}) for the case of the
hypertorus with $h=b=c=0.8$. In both calculations, we take a simple flat ($
{\cal P}(k)=1$) power spectrum with no additional physics added.

In both maps the small-scale structure and location of the spots are
identical, but in the approximate map the large-scale features of the spots
are smeared out. This smearing stems from the fact that our approximation
does not incorporate the 
damping of power at low wavenumbers due to the discretization of eigenmodes
on the compact space. Thus $C^U[d_{{\rm min}}(\vec x(\hat n),\vec x^{\prime
}(\hat n^{\prime }))]$ falls off more slowly than the corresponding
correlation function on the compact space.

\begin{figure}[tbp]
\centerline{\ \quad\quad\quad\quad\quad\quad  
\psfig{file=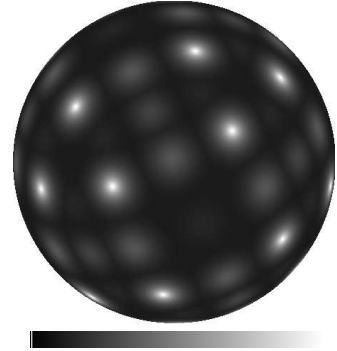,width=2.25in}} \vskip 15truept
\caption{Real-space approximated antipodal map, calculated from a flat
power-spectrum in which the dipole, quadrupole, and octopole have been
removed.}
\label{cminus4}
\end{figure}

This discrepancy can be remedied by replacing the $C^U$ with an angular
correlation function from which the lowest multipoles have been removed. In
the case of a $h=b=c=0.8$ space, the damping is well approximated by
removing the dipole, quadrupole, and octopole terms \cite{us2}. This results
in the approximate map shown in Fig.\ \ref{cminus4}, which is quite close to
the exact result. In principle, the broadening of the spots can also be
removed by including more terms in the method of images expansion, eqn. 
\ref{eq:approx}. In practice, however, a cumbersome number of terms have to
be included in order to begin to approach the exact result, with the maps
actually getting worse before they get better.

The question remains how to analyze real data so as to extract patterns with
confidence. It is customary in an analysis of CMB observations to take
angular averages in Fourier space. All standard methods thus smear out the
very patterns we seek. Instead we advocate that the correlation maps be
treated as real space pictures of geometry. As such, they are akin to
pictures of a galaxy or to observations of gravitationally lensed images.

All of the previous correlation maps are ensemble averages. Since we have
only one universe to observe and only one realization of the data we might
worry that cosmic variance would drown out any of the features of topology
in a given realization. We are unable to combat cosmic variance by averaging
over the sky since it is precisely such averaging which we are trying to
avoid. We have numerically simulated high-resolution observations of a
toroidal universe to demonstrate what future satellite observations will
offer. Cosmic variance is not a terrible hindrance and we are able to
extract the correlated information from the high-resolution simulations.

\begin{figure}[tbp]
\centerline{\quad\quad\quad\quad\quad\quad {\psfig{file=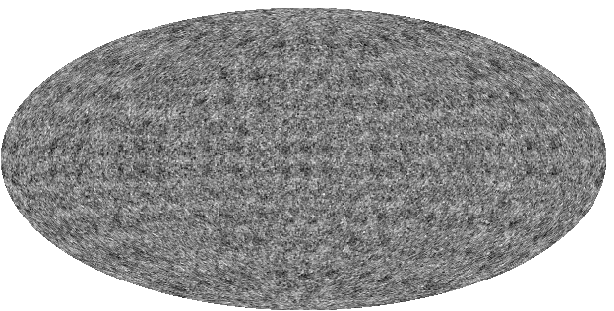,width=2in}} 
} \centerline{\quad\quad\quad\quad\quad\quad  
\psfig{file=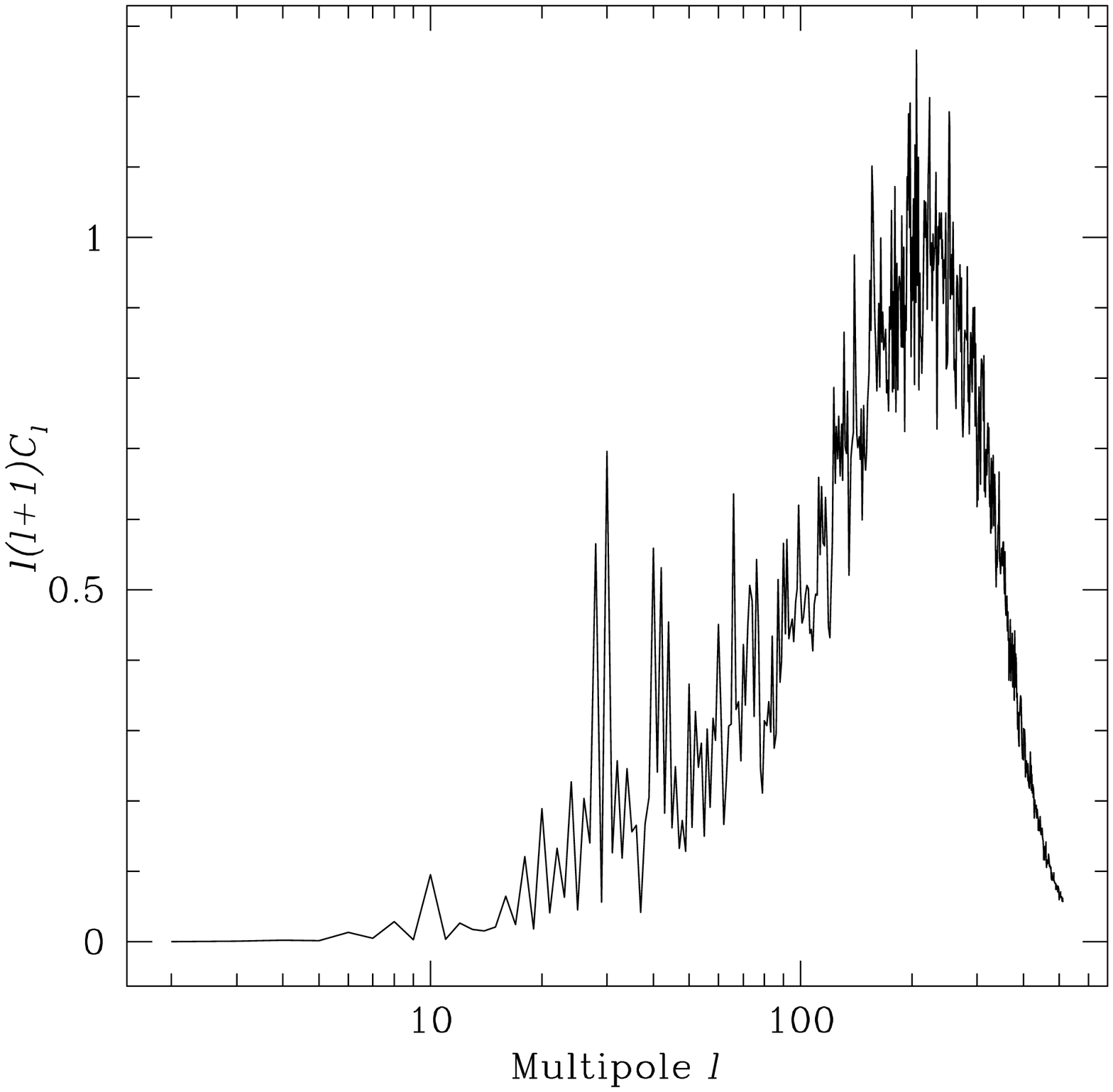,width=2.in}}
\caption{The unsmoothed numerical CMB data with 
$1024\times 513$ pixels for a small square hypertorus
with side of length $0.2\Delta \eta $. The power spectrum for this
simulation. }
\label{mapt}
\end{figure}

\begin{figure}[tbp]
\centerline{\psfig{file=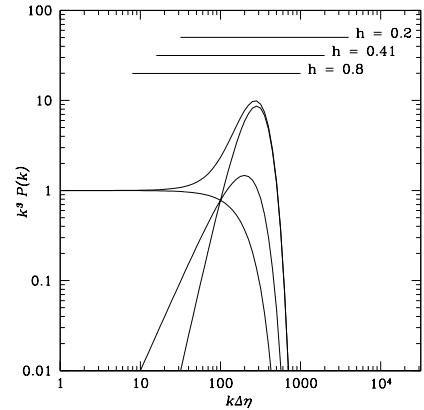,width=2.25in}}
\caption{The power spectrum as a function of $k \Delta \eta$. } 
\label{pk}
\end{figure}

The map of $\delta T(\hat n)/T$ at the top of Fig.\ \ref{mapt} is a
simulation of an all-sky map generated using the explicit eigenmodes of (\ref
{eq1}). The physical processes at work on very small scales were modeled so
as to emulate the Doppler peaks of a flat CDM universe. To obtain fairly
accurate maps we have to keep account of the physical processes at the epoch
of decoupling which determine the shape of the spectrum. At decoupling, the
comoving length of the horizon is $\eta _{dec}\simeq 10^{-2}\Delta \eta ,$
which is presumably far smaller than the dimension of our fundamental
domain. The causal processes that modulate the spectrum are therefore not
modified by the compact nature of the space. In this case we can use a
suitable function for ${\cal P}(k)$ which is able to reproduce the height and
position of the first Doppler peak in a flat CDM universe. The function we
choose is the following: 
\begin{equation}
\frac{{\cal P}(\kappa )}{\kappa ^3}=\left( \kappa ^{n-1}+\kappa ^{n+1}+\kappa
^{n+3}\right) \exp \left( -{\kappa ^2}/{4}\right)  \label{fullpk}
\end{equation}
where $\kappa =10^{-2}k\Delta \eta $ and $n$ is the primordial fluctuation
spectral index. We show this function in Fig.\ \ref{pk}. In the same figure
we also show the $k$--range covered by the model we examined. With the power
spectrum of eqn.\ (\ref{fullpk}), the relative height of the first Doppler
peak and the Sachs--Wolfe plateau are in good agreement with a flat CDM
model for $0.8\leq n\leq 1.2$. As reported in \cite{us}, the lower limit on $
k$ is fixed by the dimensions of the fundamental domain, while the upper
limit is constrained by the resolution grid in ${\bf k}$--space used to
generate a given realization. We used a grid of $n_xn_yn_z=256^3$ elements
in ${\bf k}$--space which we fast-Fourier transformed back into real space
to calculate our ${\delta T({\hat n})/T}$ spectra. The maximum wavenumber
for a given realization is $k_{max}=n_x\pi /h$. It is apparent from Fig.\ 
\ref{pk} that it is unnecessary to go beyond this resolution limit to probe
down to the accuracy of our approximate power spectrum.

To obtain the angular power spectrum, $C_\ell =\sum_m|a_{\ell m}|^2/(2\ell
+1)$, we decompose the temperature fluctuations into spherical harmonics, $
\delta T(\hat n)/T=\sum_{\ell m}a_{\ell m}Y_{\ell m}(\hat n)$, using a
modified version of the fast code developed by Muciaccia {\it et al.}\/ \cite
{natoli}. The power spectrum of the simulation can be seen at the bottom of
Fig.\ \ref{mapt}. Notice the enhancement of power due to the multiple copies
of the fundamental domain, a feature already noted in Ref. \cite{us2}. 
While the $C_\ell $'s certainly do not contain all of the information in a
map of $\delta T/T$ in a universe with multiconnected topology, the spectrum
does reveal the essential behaviour of $\delta T/T$ at very high $\ell $s
where topology is less influential.

\begin{figure}[tbp]
\centerline{{\psfig{file=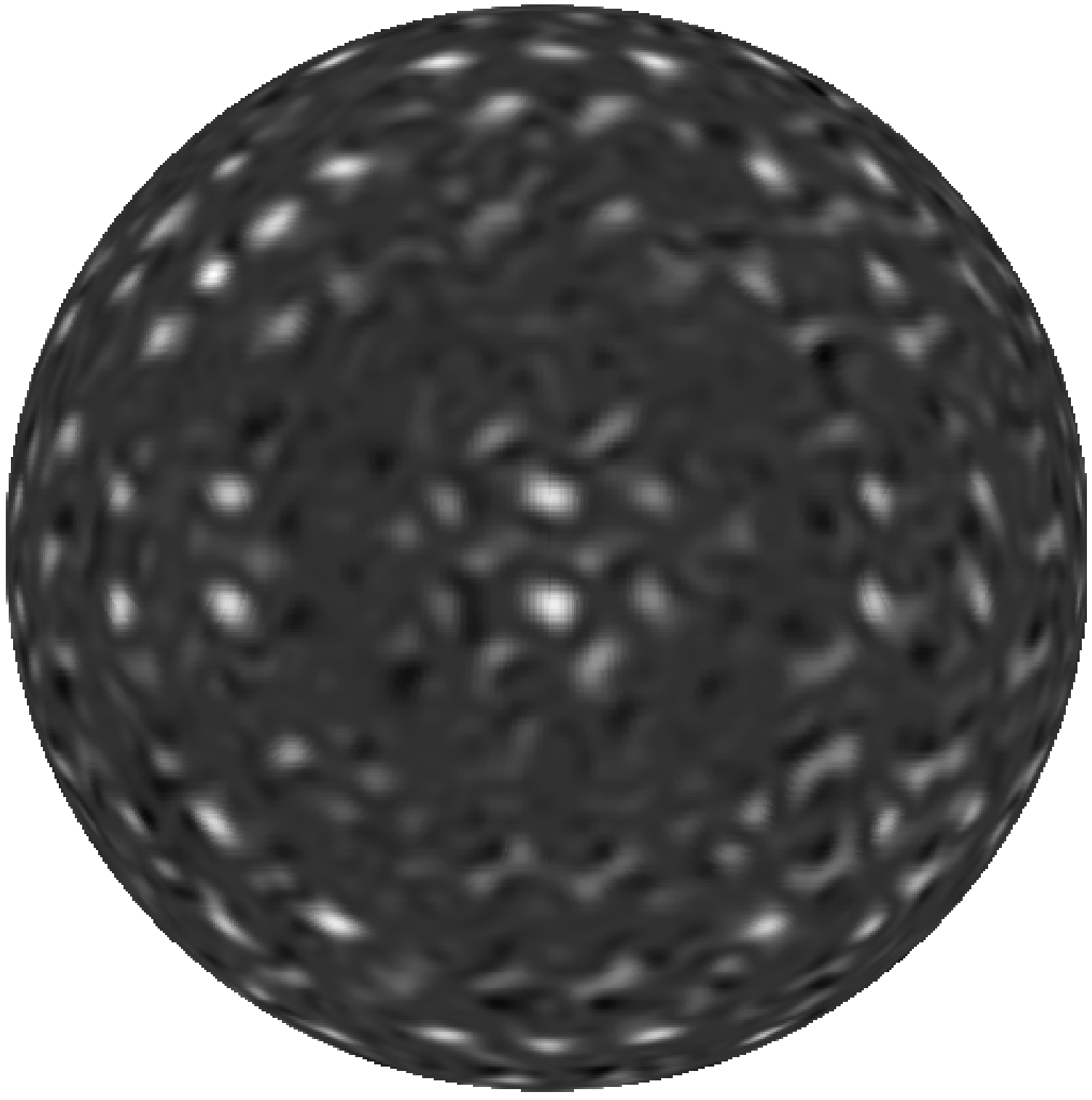,width=1.75in}} { 
\psfig{file=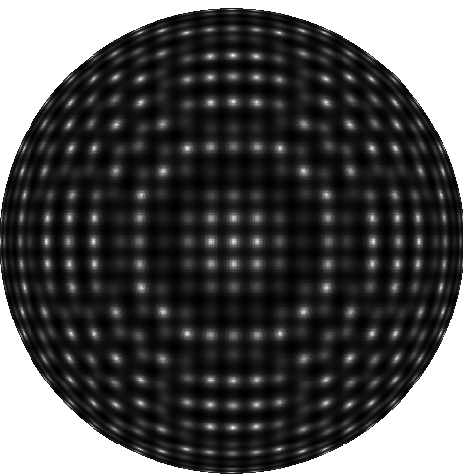,width=1.75in}}}
\caption{A comparison of the antipodal map read directly off the simulated
data shown in Fig.\ \ref{mapt} after smoothing on scales of $1.5^o$ (left)
with an ensemble average (right). }
\label{parreal}
\end{figure}

The leftmost panel in Fig.\ \ref{parreal} is the antipodal map from the
data smoothed on scales of $1.5^o$, as though the data were convolved with
an experimental beam. The ensemble average computed from eqn.\ (\ref{eq2})
is shown in the rightmost panel of Fig.\ \ref{parreal}. Clearly, the
realization shows the structure of the ensemble average. While this small
universe can be ruled out with the COBE data due to the conspicuous lack of
power on small scales \cite{{sss},{us}}, the pattern would go undetected in
the existing satellite data. If we had smoothed on scales of $7^o$ to
emulate the COBE experiment, any definitive pattern would have been washed
away. The much higher resolution of MAP and {\it Planck Surveyor} is
required to measure the universe's spots.

The very small space was chosen for dramatic effect. Since these spaces
sample low $\ell $ modes more sparsely, there is less cosmic variance in
some sense and the patterns are easier to detect. The large spaces can
suffer more contamination from low $\ell $ modes. 
To recover the bare information in the ensemble averages, the low $\ell $
modes may need to be cleaned from the larger spaces. 

\begin{figure}[tbp]
\centerline{{\psfig{file=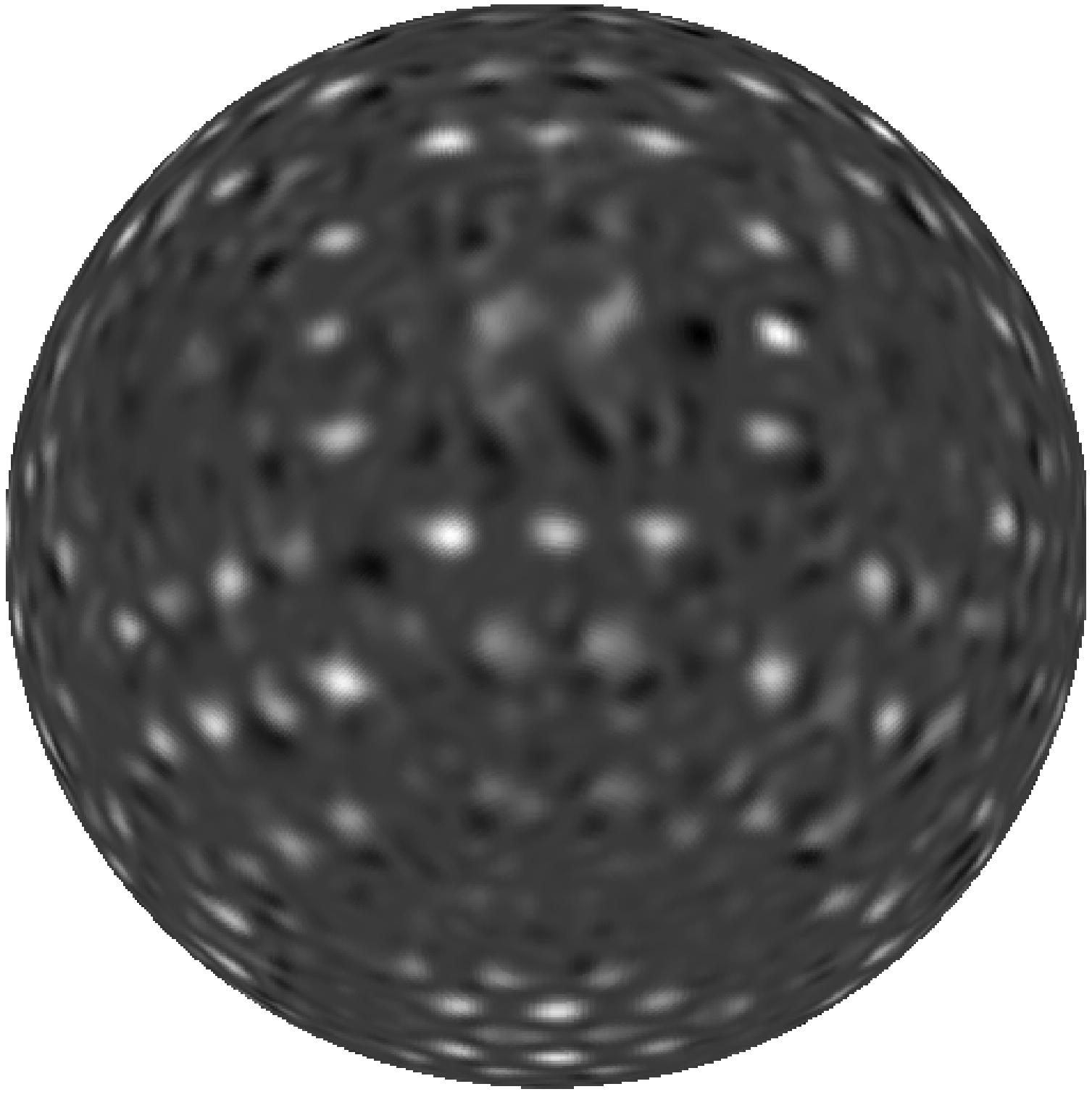,width=1.75in}} { 
\psfig{file=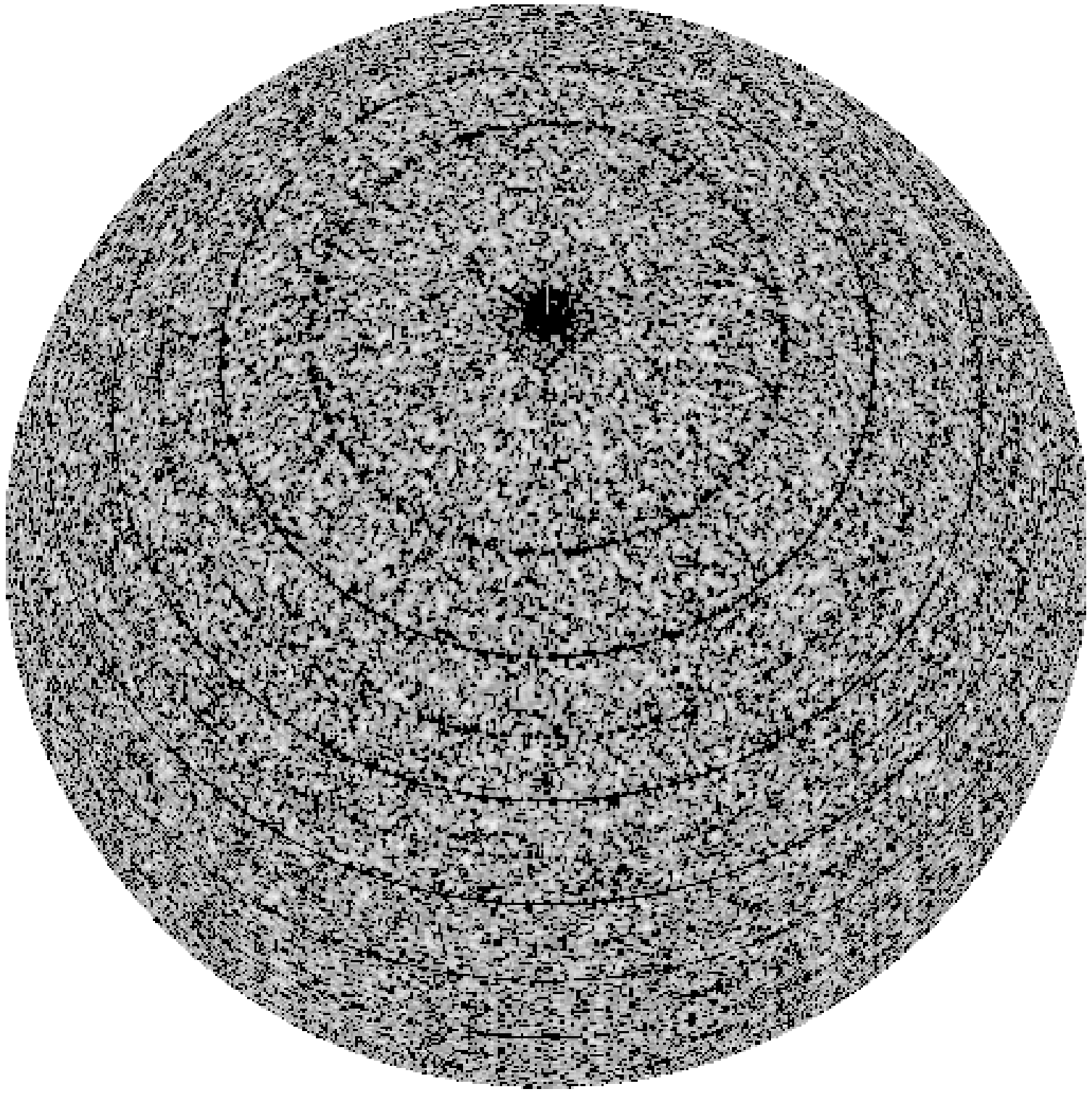,width=1.75in}}}
\caption{On the left is the correlated sphere $C_z(\hat n)$ defined in eqn. 
\ref{cz} read directly off the numerical data. On the right the quantity $
\xi_z(\hat n)$ of eqn. \ref{xiz} is plotted to draw out the circles. }
\label{cz2}
\end{figure}

We also consider the correlation which compares $\hat z\rightarrow -\hat z$,
as this is a true symmetry of the space. If we align the fundamental domain
with the $\hat z$ axis of the SLS, then the correlation compares points in
the direction $\hat n(\theta ,\phi )$ with points in the direction $\hat n 
^{\prime }(\pi -\theta ,\phi )$: 
\begin{equation}
C_z(\hat n)=C(\hat n(\theta ,\phi ),\hat n^{\prime }(\pi -\theta ,\phi )).
\label{cz}
\end{equation}
Notice that even in a universe with no topology there will be some structure
in such a correlation function. The equator so defined is always compared
with itself and so shows more correlations than the poles even without
multiconnected identifications. Since this is an actual symmetry of the
hypertorus we should find 10 circles in one hemisphere for the small space
of size $h=0.21\Delta \eta $. We pick up nine circles at latitudes $
64^o,53^o,44^o,37^o,30^o,24^o,17^o,11^o$ and $6^o$, and ten if we include
the one identified point at the poles, as shown in Fig.\ \ref{cz2} (compare
with Fig.\ \ref{hypz}). It is difficult to see that these spots do in fact
lie on fully correlated circles. After all, the maps only represent the
simple product $\delta T(\hat n)\delta T(\hat n^{\prime })/T^2$. A measure
of the correlations across the pairs of circles would draw out the feature
more crisply (see also, in this connection, the statistic suggested in Ref. 
\cite{css}).
To illustrate, we plot the quantity 
\begin{equation}
\xi _z(\hat n)=\left( {\frac{\delta T(\hat n(\theta ,\phi ))}T}-{\frac{ 
\delta T(\hat n^{\prime }(\pi -\theta ,\phi ))}T}\right) ^2\ \   \label{xiz}
\end{equation}
for the unsmoothed data, in the rightmost panel of Fig.\ \ref{cz2}. This
measure of the temperature difference singles out the circles.

\begin{figure}[tbp]
\centerline{{\psfig{file=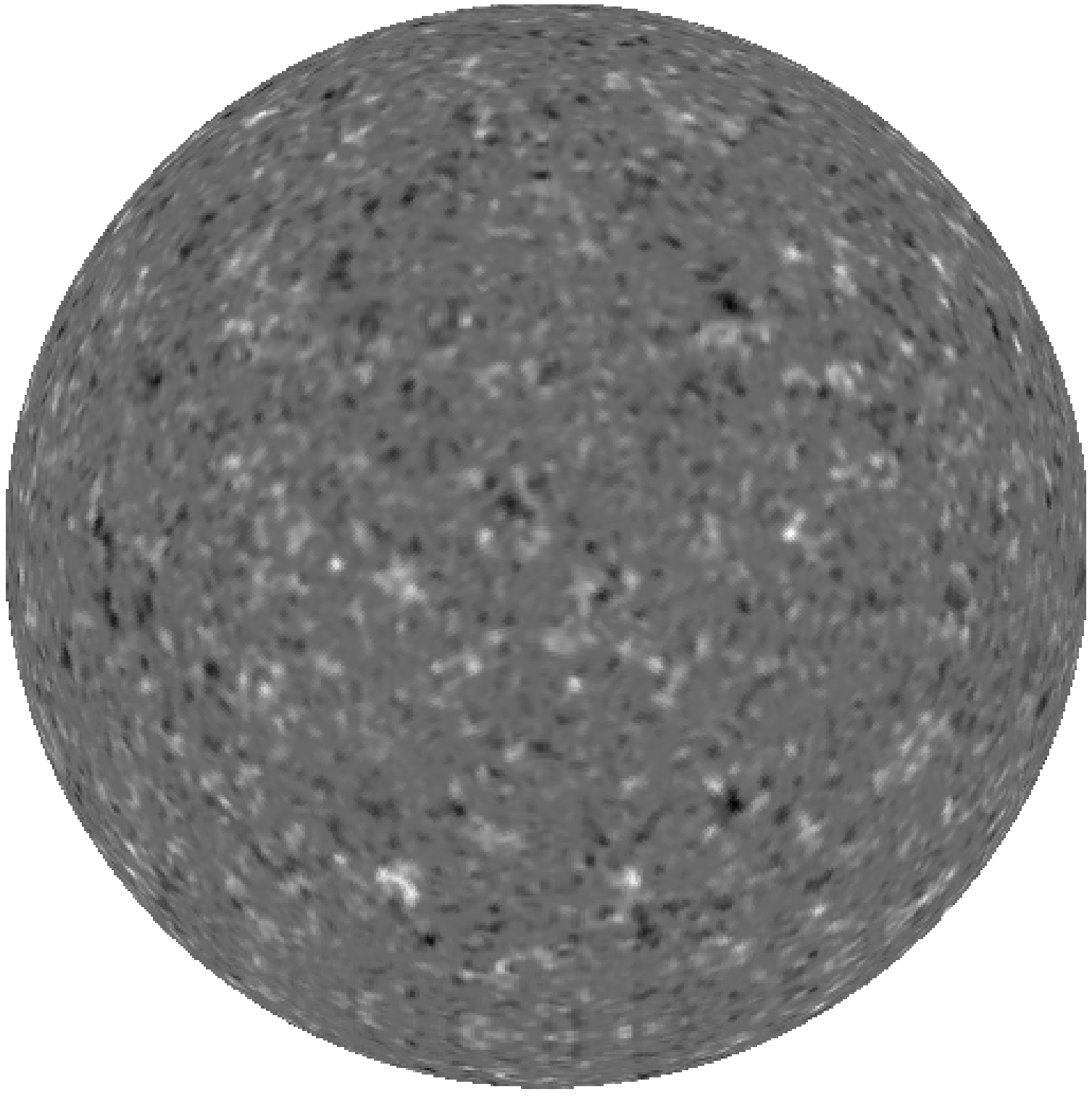,width=1.8in}} { 
\psfig{file=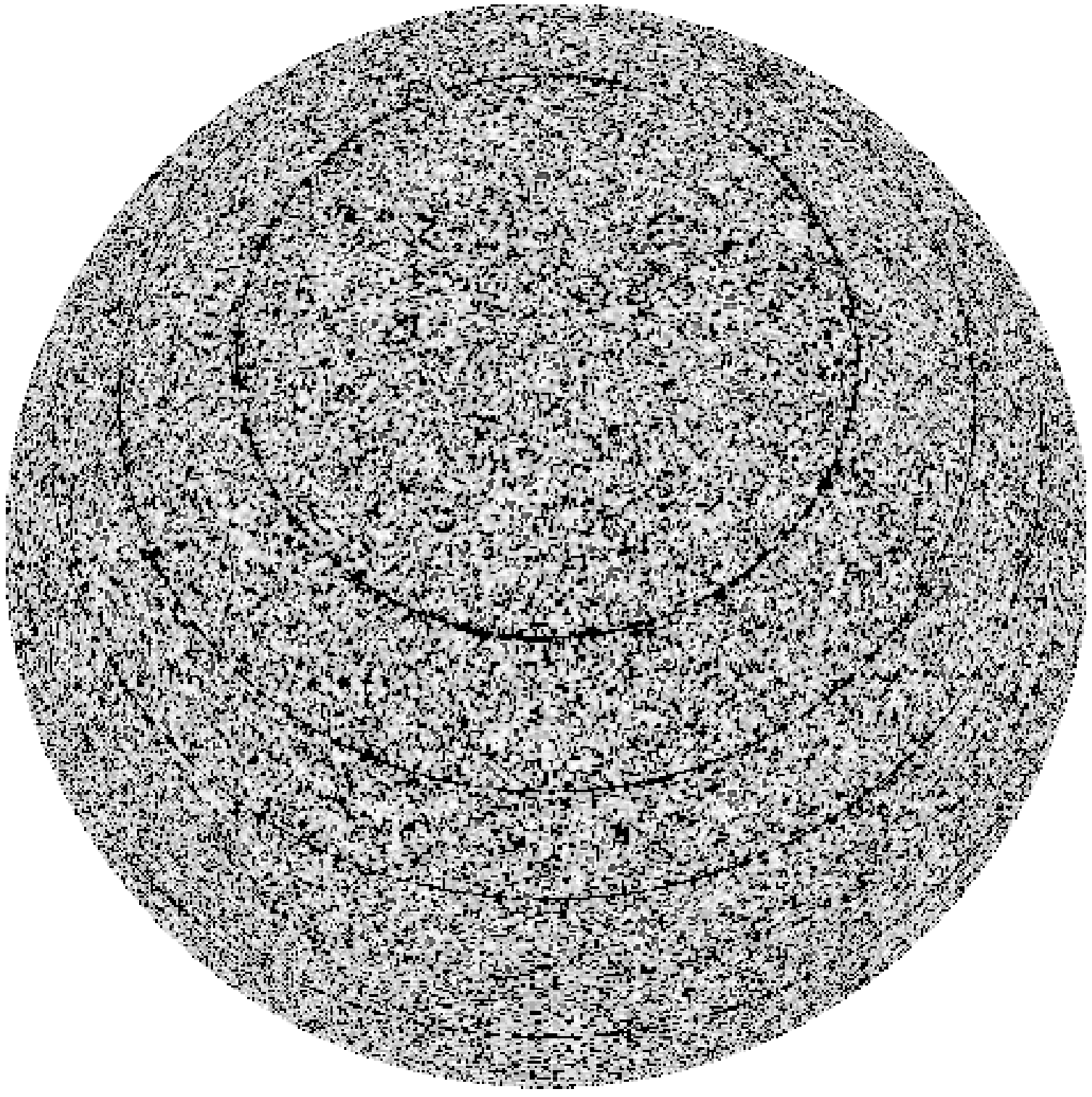,width=1.8in}}}
\caption{On the left is $C_z(\hat n)$ read directly off the numerical data
for the larger torus of size $0.41\Delta \eta$ smoothed on scales of $0.5^o$
. On the right $\xi_z(\hat n)$ of eqn. \ref{xiz} is plotted. }
\label{cz3}
\end{figure}

For a larger square torus of size $h=0.41\Delta \eta $, the correlation $C_z( 
\hat n)$ of eqn.\ (\ref{cz}) also finds faint circles. Fig.\ \ref{cz3} shows
the occurrence of circles in the sky at latitudes of around $55^o,38^o,24^o$
and $12^o$. The left figure is a map of the simple correlation function $C_z( 
\hat n)$ of eqn. (\ref{cz}) smoothed at $0.5^o$ while the right figure
locates the circles more distinctly, without smoothing, by plotting the $\xi
_z(\hat n)$ of eqn. (\ref{xiz}). The thin dark circles are the collection of
identical points for which $\xi _z=0$.

\begin{figure}[tbp]
\centerline{\psfig{file=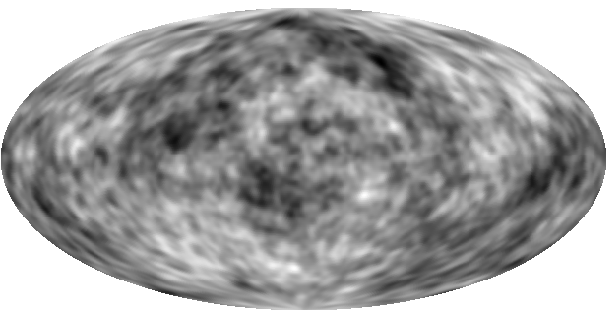,width=2.5in}} \centerline{{ 
\psfig{file=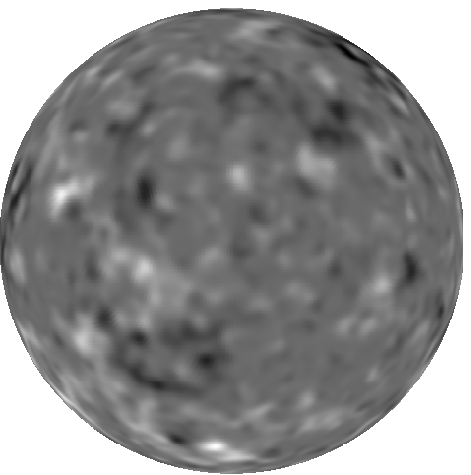,width=1.75in}} { 
\psfig{file=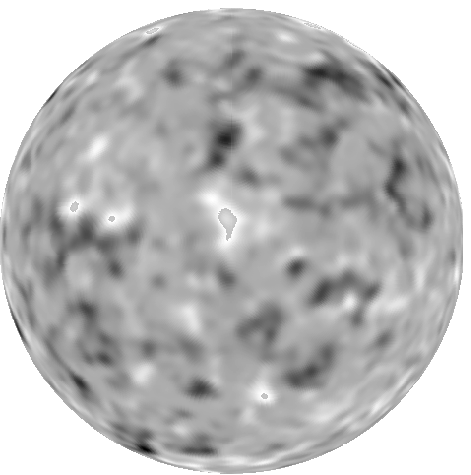,width=1.75in}}}
\caption{Simulated CDM data for a universe with no topology. The standard
map of the surface of last scatter is shown on the top. On the bottom left
is the map $A(\hat n)$ and on the bottom right is the map $C_z(\hat n)$. }
\label{notpat}
\end{figure}

For contrast, we compare the predictions for a hypertorus with those for a
simply-connected flat cosmos. Fig.\ \ref{notpat} shows a realization of the
SLS for a flat CDM universe. The data is again smoothed on scales of $1.5^o$
. The low $\ell $ modes missing from the map in Fig.\ \ref{mapt} are clearly
present in an infinite universe. There is no evidence of a pattern in
antipodal correlations, nor in the $C_z(\hat n,\hat n^{\prime }(\vec z 
\rightarrow -\vec z)),$ as demonstrated in the bland pictures of Fig.\ \ref
{notpat}.

In general we suggest a two-step investigation to observe topological
lensing as the data becomes available. Firstly, it is advantageous to smooth
the data on some small scale. 
Correlations can then be read off more easily from the smoothed map.
Secondly, the unsmoothed data can then be exploited to measure the
temperature variation across the spot at the location indicated by the
correlated signal. The measured spectra of the related spots can then be
compared. If the spectra match, topologically lensed images have been
measured: a correlation measure across two spots can be used to confirm
lensed copies of a fluctuation. The small-scale structure of the two regions
should be almost identical,\footnote{ 
The sphere of last scatter cuts through the same 3D volume differently, so
even if a point is identical to another on the SLS, the surrounding small
region can be a slightly different sample of the 3D patch.}, although
large-scale effects such as the subtraction of the CMB dipole, or the
integrated Sachs-Wolfe effect in open spaces may cause a difference in the
mean temperature of the two patches. In practice, a generalization of eqn.\ ( 
\ref{xiz}) such as 
\begin{equation}
\xi (\vec x)=\left( {\frac{\delta T(\vec x-\vec x_{{\rm spot}})}T}-{\frac{ 
\delta T(\vec y-\vec y_{{\rm spot}})}T}\right) ^2  \label{xix}
\end{equation}
where the temperature fluctuations are defined with respect to the local
mean temperatures of the spots, should do well in confirming lensed copies
of fluctuations.

As we improve our understanding of the correlation function on short scales
we might determine the spectrum of each of these spots and use this
information to distinguish fictitious correlations from the real thing. In
an actual realization, all of these correlated spots will be there under a
web of spurious correlations due to cosmic variance. These random
correlations can be distinguished much as foreground sources are
distinguished from the galaxies they occult. Spurious correlations will not
be distributed on rings, nor will they have the characteristic size and
spectrum of the topological correlations. Cosmic variance as a form of
cosmic noise could thus in principle be subtracted off the maps. Again, the
task is similar in spirit to distinguishing the gravitationally lensed
images of a quasar from other unrelated or foreground sources.

\vfill\eject

\section{The Hyperbolic Zoo}

\label{zooh}

We turn now to the application of our correlated spheres to compact
hyperbolic universes. Compact hyperbolic spaces are inherently chaotic. The
exponentially deviating trajectories of geodesic motions on a space of
negative curvature mix and fold chaotically through the space as they exit
and enter the multi-faceted fundamental domain. Chaos endows these spaces
with many intriguing properties, including fractal structures within the pattern
of entangled geodesics. The patterns inscribed in these skies thus promise
to be intricate.  
Primordial quantum fluctuations which ultimately seed the hot and cold spots
are described by quantum chaos for which there are very different
predictions than for nonchaotic quantum systems \cite{gutz}. The assumption
of a flat, Gaussian seeded spectrum of perturbations
may be a poor one.

While complicated, these spaces are not obscure. There is a countable
infinity of topologically distinct compact ${\bf H}^3$ spaces, although they
have yet to be completely classified \cite{thur}. 
Furthermore, observations favor a universe with subcritical density. There
is therefore considerable interest in understanding the predictions for
these spaces. 
Regardless of how predictive maps are produced, the ultimate question is,
how do we analyze the CMB data to search for topology. Canonical treatments
rely on angular averages which smear out patterns and a likelihood analysis
based on Gaussian statistics. While this may have some restricted meaning,
it is dangerous to draw precise conclusions from a Gaussian isotropic
probability distribution when the space itself destroys isotropy and the
primordial spectrum is unknown. 
Furthermore, it requires a case by case analysis and may even depend on the
location of the observer.

This implies that a statistical analysis of the data requires a model
template. 
If the universe is not a perfect  manifold of constant curvature the
template match is lost. Instead of asking the statistical fit of the data to
an infinite number of models we can just take a picture of the sky and from
this obtain a picture of correlated maps. As already argued, the spectrum of
the spots can be measured to judge if we are really looking at the
topological lensing of the horizon at the time of decoupling. 
The idea of combing the data for circles in the sky also shares the
model independence feature and motivation.

We apply our method to two small hyperbolic topologies, the Weeks space and
the Best space, named after their discoverers. The Best space is a compact
hyperbolic manifold obtained by identifying the twenty faces of a regular
icosahedron \cite{best}. The Weeks space \cite{weekspc} has a more
complicated fundamental domain with 18 faces and is of particular interest
since it is currently the smallest compact hyperbolic space known.

\subsection{Strong patterns in a Weeks space}

\begin{figure}[tbp]
\centerline{{\psfig{file=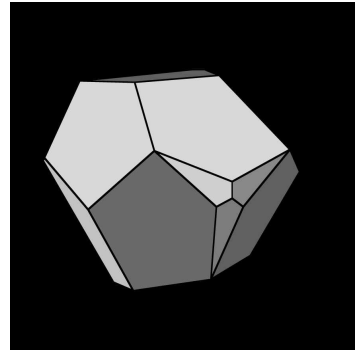,width=1.8in}}}
\caption{The Dirichlet domain for the Weeks space. }
\label{weekpic}
\end{figure}

It is advantageous to consider ${\bf H}^3$ embedded as a $3D$ surface in a $
4D$ Minkowski spacetime. The universal cover ${\bf H}^3$ is then a
pseudosphere and the $4D$ coordinates are restricted to the $3D$ surface
with pseudoradius $-1$, 
\begin{equation}
-u_o^2+u_1^2+u_2^2+u_3^2=-1\ \ .
\end{equation}
The isometries of ${\bf H}^3$ can then be written as $4\times 4$ matrices of
the special Lorentz transformations. The coordinate transformation 
\begin{eqnarray}
u_o &=&\cosh r  \nonumber \\
u_1 &=&\sinh r\sin \theta \cos \phi   \nonumber \\
u_2 &=&\sinh r\sin \theta \sin \phi   \nonumber \\
u_3 &=&\sinh r\cos \theta 
\end{eqnarray}
recovers the cosmologically familiar form of the $3D$ metric distance in
comoving coordinates, 
\begin{equation}
ds^2=dr^2+\sinh ^2r(d\theta ^2+\sin ^2\theta d\phi ^2).
\end{equation}

The geodesics take on a particularly simple form \cite{bl} in the
Minkowskian space. For simplicity, we tended to leave the Earth at the
origin. For completeness, we mention how to move the Earth away from the
center of the universe. If we align the Earth with the $z$-axis at position $
u_e^\mu =(u_e^o,0,0,u_e^3)$ we can Lorentz boost the Earth to the origin
with the transformation 
\begin{equation}
\Lambda =\pmatrix{\gamma &0&0&-\gamma \beta \cr 0&1&0&0\cr 0&0&1&0\cr
-\gamma \beta &0&0&\gamma }
\end{equation}
and $\gamma \equiv u_e^o$ and $\gamma \beta \equiv u_e^3$. A photon observed
in the direction $\hat n$ has $\hat v=-\hat n$ and thus originated on the
SLS at coordinates 
\begin{equation}
u=\Lambda ^{-1}\pmatrix{\cosh \Delta \eta \cr \hat n \sinh \Delta \eta }\ \ .
\end{equation}
This can easily be generalized to an arbitrary location with an arbitrary
Lorentz boost.

The radius of the SLS, $\Delta \eta $, depends on the value of $\Omega _o$
and the redshift of last scattering. In general, 
\begin{equation}
\eta ={\rm arccosh}\left( 1+{\frac{2-2\Omega _o}{\Omega _o(1+z)}}\right) 
\end{equation}
in units of the curvature radius. The volume of the SLS is 
\begin{eqnarray}
V_{SLS} &=&\int \sinh ^2rdrd\Omega   \nonumber \\
&=&\pi (\sinh (2\Delta \eta )-2\Delta \eta )\ \ .
\end{eqnarray}
The volume of the SLS grows exponentially with time. Therefore, many more
copies of the fundamental domain can be contained within a surface of last
scatter. The number of copies is a topological invariant, quite unlike in
flat space. In flat space, one is free to set the volume of the manifold
relative to the volume of the SLS arbitrarily. This is not possible for $
{\bf H}^3,$ as is ensured by the rigidity theorem \cite{rigid}, which states
that the volume in units of the curvature is fixed. A peculiar consequence
is that the volume of a manifold is a topological invariant. In effect, if
we measure topology we can refine our measure of the radius of the last
scattering surface and thereby $\Omega _o$. In other words, we can use
topology to measure curvature.

In order to compactify the space, we consider the specific example provided
by the Weeks space. The fundamental domain is a polyhedron with 18 faces and
26 vertices shown in Fig.\ \ref{weekpic} which was taken from {\it SnapPea},
a census of compact hyperbolic manifolds \cite{snap}. The Weeks space is the
smallest 3-manifold known with a volume of $\simeq 0.94$. With $\Omega _o=0.3
$, and the redshift of last scattering taken to be $z=1100$, so $\Delta \eta
=2.328$, we get $V_{SLS}=150.64,$ and so there are roughly $150$ copies of
this universe within the SLS. With $\Omega _o\simeq 0.6$, $\Delta \eta \sim
1.5$ and there are only about $5-6$ copies.

There are 9 identification rules to glue these 18 faces in pairs. The 9 $g_i$
can be used to define a set of generators with many relations among them
(such a set of generators was also used by Fagundes \cite{fagundes}). They
are related to the words ${a,b}$ of a much simpler presentation of the
fundamental group $\{a,b:ababa^{-1}bba^{-1}b,abab^{-1}aab^{-1}ab\}$. We
extract from {\it SnapPea} these nine face-pairing $O(3,1)$ matrices and
their inverses out to 12 significant digits. It is necessary to use
extremely precise matrices due to the chaotic flows. The sensitivity to
initial conditions causes the image points to quickly be lifted off the
pseudosphere if insufficient precision is used.

The image points are denoted by $y^\mu $ in Minkowski coordinates: 
\[
y_{k_n,...,k_1}^\mu =\prod_i^ng_{k_i}u^{\prime \mu }
\]
and $k_i=\pm 1,..,\pm 9$. Since we number the $g_k$ from $1-9$, let $g_o$ be
the identity so that $y_0^\mu \equiv {u}^{\prime \mu }$. We want to find the
closest separation between $u^\mu $ and any of the image points $y^\mu $.
The $g_{k_i}...g_{k_n}$ form words of length $n$ where $n$ is the highest
order of the farthest neighbor.

\begin{figure}[tbp]
\centerline{\quad\quad\quad\quad\quad\quad { 
\psfig{file=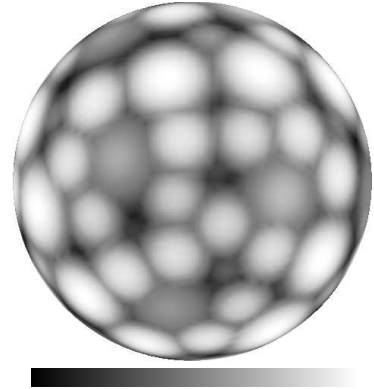,width=2.5in}}} \vskip 15truept
\caption{Correlation with a random point and the rest of the sphere for the
Weeks space with $\Delta \eta=1.5$. }
\label{weekscorr}
\end{figure}

\begin{figure}[tbp]
\centerline{\quad\quad\quad\quad\quad\quad { 
\psfig{file=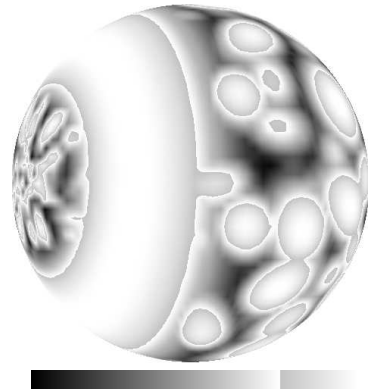,width=2.5in}}} \vskip 15truept
\caption{The sphere correlated under the transformation $g_8$. }
\label{weeksg1}
\end{figure}

Another way to count copies is to count periodic geodesics. The
identification rules correspond to the minimal closed loop geodesics. The
number of geodesics of length $L$ goes as 
\[
N(L)\sim {\frac{e^{hL}}{hL}}
\]
which is a result from chaos theory, with the Kolmogorov-Sinai entropy
determined by the volume scale, $h\sim V^{-1/3}$. If we assume crudely that $
L=nV^{1/3}$ so that $n$ is the order of the neighbor, that is, the length of
the word $y^\mu $, then 
\[
N(n)\sim {e^n/n}
\]
So, for $\Omega _o=0.3$, the SLS encompasses roughly $150$ copies of the
fundamental domain and includes copies which are between $6$ and $7$ words
away. For $\Omega _o=0.6$, the farthest neighbors are about $3$ words from
the origin. The $9$ generators and their inverses combine to form $18^n$
words of length $n$ but as a result of the many relations among the $g_k$,
all but $N(n)\sim e^n/n$ of these are repeats. Scanning the $18^n$ possible
images $N_{pix}\times N_{pix}$ times is a huge numerical demand. To manage
the task we use the following numerical algorithm, which is analogous the
scheme used in the flat cases:

(1) Move the point $u$ on the SLS in towards the origin two steps to the
image points $y_{k_2k_1}^\mu =g_{k_2}g_{k_1}u^\mu $. Out of these $18^2$
points, select the three nearest the origin.

(2) Move these 3 images two more steps and keep the one image nearest the
origin so that it is essentially within the fundamental domain. Depending on
how many neighbors fit within the SLS, step (2) can be repeated.

(3) Repeat the same steps to the point $u^{\prime}$ being compared. Both
points should now be located within the fundamental domain.

(4) Keeping one image fixed within the fundamental domain, move the other
within two neighbors until the distance between the two images is minimized.

(5) The geodesic distance on the pseudosphere is 
\begin{equation}
d(y^\mu ,y^{\prime \mu })={\rm arccosh}\left( y^oy^{\prime o}-\vec y\cdot 
\vec y^{\prime }\right) .\ \ 
\end{equation}
We use this in our approximation 
\begin{equation}
C_{{\cal M}}(\hat n,\hat n^{\prime })\approx C^U(d_{{\rm min}}(\vec x(\hat n 
),\vec x^{\prime }(\hat n^{\prime }))),\ \   \label{repeat}
\end{equation}
where we use the angular correlations obtained from the CMBFAST code, with $
\Omega _b=0.05$ and the remainder of the subcritical density is assumed to
be made up of dark matter.

It is important to note that in a negatively-curved cosmos there is, in
addition to the Sachs-Wolfe effect on the surface of last scattering, an
integrated Sachs-Wolfe (ISW) contribution to the perturbations. The ISW
effect is due to the decay of the gravitational potential as the photons
transit the space. Although we have not yet fully included the ISW effect,
only the fluctuations on the largest scales should be affected whereas spots
probe small-scale physics. On those scales where the ISW effect contributes
the geodesics are deviating sufficiently so that photons that originated in
the same vicinity quickly take different paths with different decaying
gravitational potentials. Correlations will not therefore be enhanced. One
might fear that topological correlations could be erased by the different
histories of the two initially adjacent photon trajectories. Since the ISW
effect does not effect the Doppler peaks in an infinite cosmos, it should
leave the universe's spots unmarred. The next phase of investigation should
include the ISW effect. 
Our maps correspond to data for which the lowest multipoles have been
cleaned off.

The first correlation we compute compares a random point with the rest of
the surface of last scattering, $C(u^\mu ,\hat n)$, as shown in Fig.\ \ref
{weekscorr}. This is unlike any correlation we have considered so far, but
it can clearly be quite successful at uncovering geometric properties. As in
the flat universes, the spots are likely spread out since we have not
accounted for the discretization of the harmonics of the finite box which
causes big dips in power at large modes, especially for such a small space.
Again, the inclusion of a huge number of terms would be needed to
incorporate this effect. 
We also compute $C(\hat n,g_8\hat n/||g_8\hat n||)$ in Fig.\ \ref{weeksg1}.
This transformation combines a boost along the hyperbolic surface with a
rotation.

\subsection{The Best patterns}

\begin{figure}[tbp]
\centerline{{\psfig{file=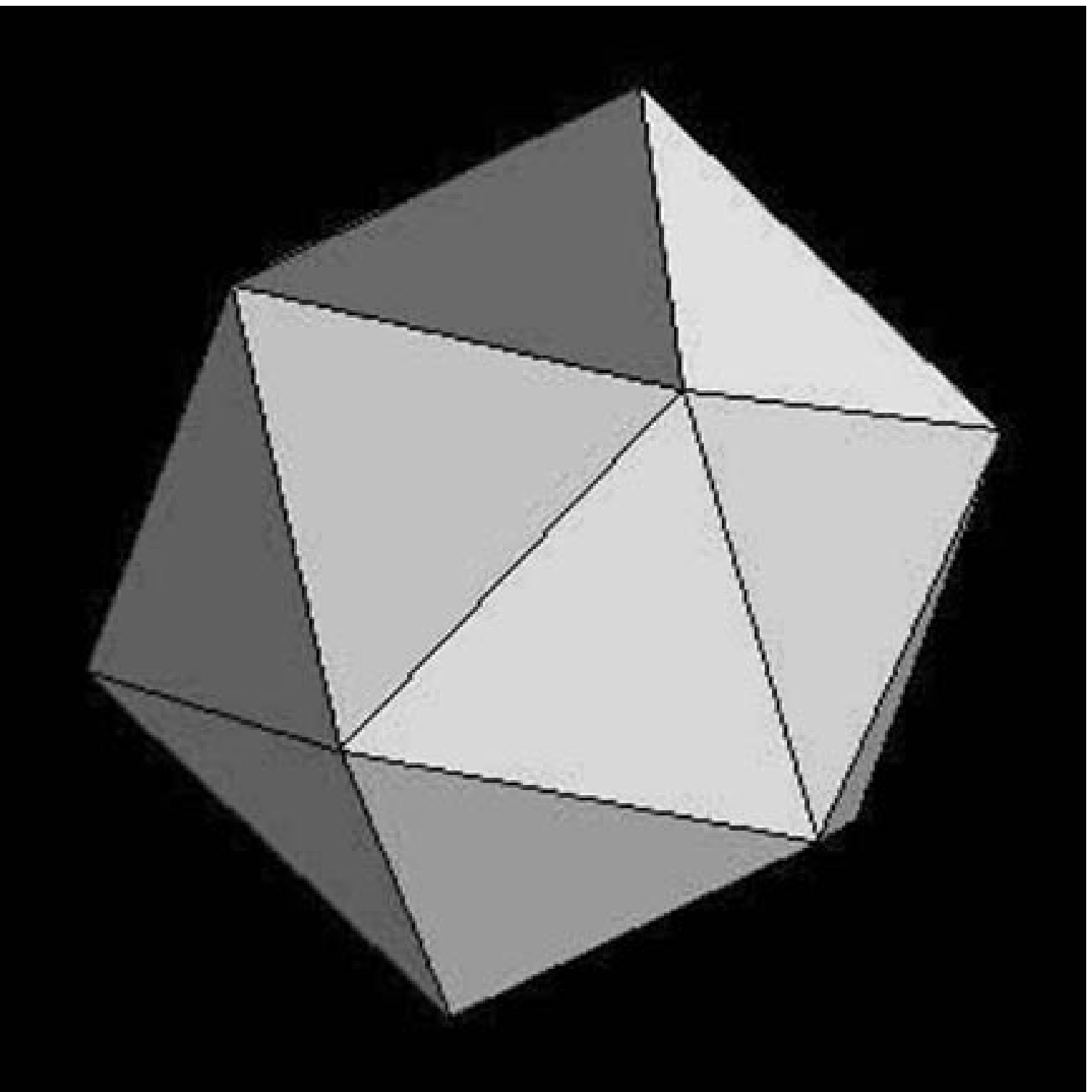,width=1.8in}}}
\caption{}
\label{bestpic}
\end{figure}

Best built three compact hyperbolic spaces by identifying the faces of an
icosahedron possessing twenty triangular faces, as shown in Fig. \ref
{bestpic}. All of these spaces have volume $V=4.6860342738$. The smallest
geodesic ball which can encompass this space has a radius of radius $r\simeq
1.38$ while the largest sphere which can be inscribed within the icosahedron
has radius $r\simeq 0.87.$ For a rough estimate of the number of copies we
count $N=V_{SLS}/V$. With $\Omega _o=0.3,$ there are on the order of $32$
copies of this Best space within the SLS.

\begin{figure}[tbp]
\centerline{\quad\quad\quad\quad\quad\quad { 
\psfig{file=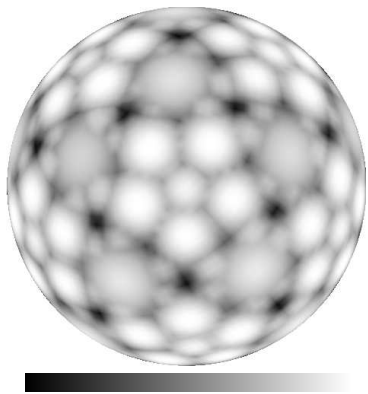,width=2.5in}}} \vskip 15truept
\caption{The correlation of one point on the surface of last scatter with
the rest of the sphere. The point is near the origin of one of the clones of
the fundamental domain. }
\label{bestpat}
\end{figure}

\begin{figure}[tbp]
\centerline{\ {\psfig{file=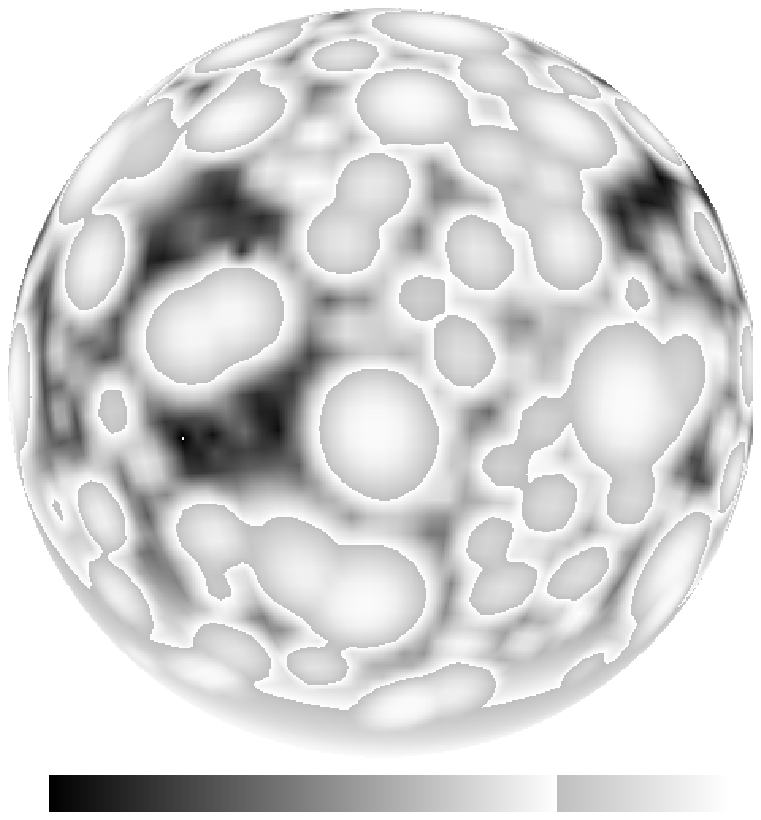,width=1.8in}} { 
\psfig{file=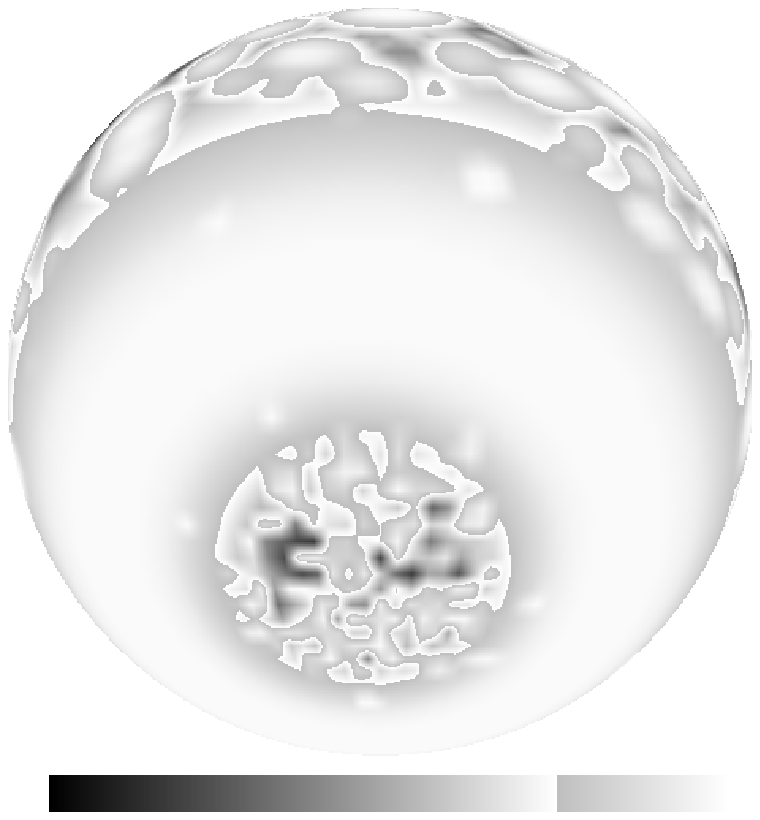,width=1.8in}}} \vskip 15truept
\caption{The sphere correlated by the $g_6$ transformation. }
\label{bestg6}
\end{figure}

The faces are identified with $10$ generators, the matrix representations
for which can again be found in {\it SnapPea}. 
The abelianized homology group is ${\bf Z/35}$. Fagundes also studied this
space and, in particular, the occurrence of periodic quasar images \cite
{fbest}. The other three non-isomorphic Best spaces with icosahedra as
fundamental domains have different fundamental groups and homology groups of 
${\bf Z/29}$ and ${\bf Z/2}\times {\bf Z/2}$. To construct our maps, we
follow the same procedure as detailed for the Weeks space. The cosmic soccer
ball in Fig.\ \ref{bestpat} is the correlation of one point on the SLS with
the rest of the sphere. The point happens to be very near the origin of one
of the copies of the icosahedron. The patterns in the plot reflect the
extreme symmetry of the fundamental domain and also hint at the fractal
nature of the geodesics. Notice the five-pointed star surrounding the
tetrahedron. In the triangular corners of the five-pointed star there appear
to be six-pointed stars surrounding hexagons. We also show an example of the
correlation $C(\hat n, g_6\hat n/||g_6\hat n||)$ in Fig.\ \ref{bestg6}.

As data from the planned satellite missions becomes available, the CMB can
be scanned for any hidden geometric features. In the meantime, these
correlated spheres show the huge potential for a pattern-oriented search of
topological lensing.

Our real-space approximation allows one to calculate 
temperature correlations while avoiding the analytically intractable
eigenvalue problem on compact hyperbolic 3-spaces. Once a candidate compact
universe has been established, such templates are useful for deeper
statistical studies \cite{bps}. However, an observational search for
topology through large-angle temperature correlations can be made without a
particular template being presupposed. This model independence of a
pattern-driven approach is particularly important because it allows us to
take pictures even if the space does not have constant curvature, if
observations depend strongly on the location of the 
observer, or if chaotic mixing leads to unusual
primordial spectra. Since we cannot predict topology within existing
theories, we need this flexibility.

While general relativity predicts the evolution of curvature, it does not
specify the topology of space or of spacetime. Only a theory beyond
Einstein's will be able to fully specify the geometry of the universe.
Supergravity theories necessarily acknowledge the importance of topology and
compact hyperbolic cosmologies have even recently been studied as a
consequence of string theories \cite{horo}. However the universe was born,
it was endowed with some topology and will have a place in our cosmic zoo.
Astronomical studies
of the topology of the universe may provide the most important 
insights into
those aspects of the fundamental laws of Nature that dictated the global
character of the space in which we live.

Geometry effects not just the large-scale universe,
but also the animals which inhabit it.
As progeny of the universe, animals inherit certain cosmic
blueprints.   Perhaps it is only fitting that
many analogues to the cosmic patterns could be found
here on Earth on the backs of insects, in animal markings, even in human 
made monuments.
If we could create a zoo of universes, each with a different 
topology, we might replicate all the animal markings from 
zebra stripes to leopard spots.

\begin{center}
\_\_\_\_\_\_\_\_\_\_\_\_\_\_\_\_\_\_\_\_\_\_\_\_\_\_\_
\end{center}

\vskip10truept

We would like to thank our many colleagues A. Balbi, J.R. Bond, N. Cornish,
P. Ferreira, K. Gorski, P.Natoli, D. Pogosyan, T. Souradeep, D. Spergel, G.
Starkman,
and J. Weeks for their invaluable contributions. This research has been
supported in part by grants from the NSF, DOE and the PPARC of the UK.

\end{document}